\documentclass[dvipdfmx,usenames]{ptephy}
\usepackage{boites}
\usepackage{latexsym}
\usepackage{amssymb}
\usepackage{amsfonts}
\usepackage{amsmath}
\usepackage{bm}
\usepackage{graphicx}
\usepackage{color}
\usepackage{ulem}
\usepackage{multirow}
\allowdisplaybreaks
\renewcommand{\a}{\alpha}
\renewcommand{\b}{\beta}
\renewcommand{\c}{\gamma}

\newcommand{\e}{\epsilon}

\newcommand{\m}{\mu}
\newcommand{\n}{\nu}
\renewcommand{\t}{\tau}
\newcommand{\z}{\omega}
\newcommand{\G}{\Gamma}
\newcommand{\C}[1]{\mathcal{#1}}
\newcommand{\T}[1]{\text{#1}}
\Year{2015}
\artid{033E01}
\DOI{ptv012}
\begin{document}

\title{Gravitational waves from a particle in circular orbits around a rotating black hole to the 11th post-Newtonian order} 

\author{\name{\fname{Ryuichi} \surname{Fujita}}{1,2}}

\address{\affil{1}{
CENTRA, Departamento de F\'{\i}sica, Instituto Superior T\'ecnico, 
Universidade de Lisboa, Av. Rovisco Pais 1, 1049 Lisboa, Portugal 
}
\affil{2}{
Departament de F\'isica, Universitat de les Illes Balears, 
Cra.\ Valldemossa Km.\ 7.5, Palma de Mallorca, E-07122 Spain
}
\email{ryuichi.fujita@ist.utl.pt}}

\begin{abstract}
We compute the energy flux of the gravitational waves radiated by 
a particle of mass $\m$ in circular orbits around a rotating black hole 
of mass $M$ up to 
the 11th post-Newtonian order (11PN), i.e. $v^{22}$ beyond the leading 
Newtonian approximation where $v$ is the orbital velocity of the particle. 
By comparing the PN results for the energy flux with high-precision 
numerical results in black hole perturbation theory, 
we find the region of validity in the PN approximation becomes larger 
with increasing PN order. 
If one requires the relative error of the energy flux 
in the PN approximation to be less than $10^{-5}$, the energy flux 
at 11PN (4PN) can be used for $v\lessapprox 0.33$ ($v\lessapprox 0.13$). 
The region of validity can be further extended to $v\lessapprox 0.4$ 
if one applies a resummation method to the energy flux at 11PN. 
We then compare the orbital phase during a two-year inspiral 
from the PN results with the high-precision numerical results. 
We find that for late (early) inspirals when $q\le 0.3$ ($q\le 0.9$), 
where $q$ is the dimensionless spin parameter of the black hole, 
the difference in the phase is less than 1 ($10^{-4}$) rad and hence 
these inspirals may be detected in the data analysis for space detectors 
such as eLISA/NGO by the PN templates. 
We also compute the energy flux radiated into the event horizon for a particle 
in circular orbits around a non-rotating black hole at 22.5PN, i.e. 
$v^{45}$ beyond the leading Newtonian approximation, which is comparable 
to the PN order derived in our previous work for the energy flux 
to infinity at 22PN. 
\end{abstract}

\subjectindex{E01, E02, E20, E31, E36} 

\maketitle
\section{Introduction}
\label{sec:intro}

Extreme mass ratio inspirals (EMRIs) are among the main candidate 
sources of gravitational waves (GWs) for future space-based detectors, 
such as eLISA~\cite{eLISA}. In EMRIs, a stellar-mass compact object 
of mass $\m$ orbits around a super-massive black hole of mass $M$. 
Due to the loss of the energy and the angular momentum by 
the emission of gravitational waves, the compact object spirals into 
the super-massive black hole. A conventional method to detect 
gravitational waves and to extract the physical information on the sources 
is matched filtering, which correlates the template bank of 
theoretical waveforms of GWs with the noisy data stream of the detector. 
In order to avoid significant dephasing in the matched filtering, 
we need to prepare theoretical waveforms whose accuracy is at least 
one part in $10^{5}-10^{6}$ since 
the accumulated phase of gravitational waves from EMRIs 
during mission time for future space-based detectors, $\sim $ yr, 
is millions of radians. 

Since the mass ratio is very small $\m/M\ll 1$, 
EMRIs can be described by the black hole perturbation theory in which 
the mass ratio is used as an expansion parameter~\cite{chapter,ST}. 
To the lowest order in the mass ratio, the small object moves on 
a geodesic of the black hole spacetime. To the first order in the mass ratio, 
the orbit deviates from the geodesic because of the 
gravitational self-force~\cite{Barack2009,PPV2011,Thornburg2011}. 
In the black hole perturbation theory, one may accurately compute 
the gravitational waves and the self-force in the strong field since 
there is no assumption on the velocity of the small object. 
However, costs for numerical calculations are so high that 
one cannot perform calculations for all the parameter space of EMRIs 
with sufficient accuracy~\cite{Gair2004}. 
Thus, from the point of view of computational cost 
it is useful if there are analytic methods to investigate 
gravitational waves from EMRIs. 

The post-Newtonian (PN) approximation to the Einstein equations is 
a standard method to compute gravitational waveforms from 
inspiraling compact binaries~\cite{post_Newton}. 
In the PN approximation to the compact binary system, one assumes that the 
velocities of the binary are much smaller than the speed of light, $v/c\ll 1$. 
In the standard PN approximation, 
the amplitude of gravitational waves and the orbital phase 
are, respectively, derived up to 3PN and 3.5PN, i.e. $v^6$ and $v^7$ 
beyond the leading order for the non-spinning compact binaries in 
quasi-circular orbits~\cite{DJS01,BDE04,BFIJ02,BDEI04,K07,BFIS08,Favata09}. 
(Note that the 3.5PN amplitudes for $(\ell,m)=(2,2),(3,3)$ and $(3,1)$ modes 
are derived in Refs.~\cite{FMBI2012,FBI2014}.) 
For the case of the spinning compact binaries in quasi-circular orbits, 
spin-orbit effects in the orbital phase are derived up to
4PN~\cite{MBBB2014}. Spin-spin effects in the orbital phase are derived 
up to 2PN~\cite{Gergely1999,Gergely2000,BVG2005,RBK2009}. 

Using the PN approximation in the black hole perturbation theory, 
high PN order expressions for gravitational waves can be obtained 
more systematically than using the standard PN approximation~\cite{ST}. 
The energy flux to infinity up to 5.5PN (4PN) for the case of a test particle 
in circular orbits around a Schwarzschild (Kerr) black hole was derived 
in Ref.~\cite{TTS} (Ref.~\cite{TSTS}) by solving 
the Teukolsky equation~\cite{Teukolsky1973}, which is 
the fundamental equation in the black hole perturbation theory. 
More recently, very high PN order expressions in the energy flux to infinity 
and gravitational waveforms for a test particle in circular orbits around 
a Schwarzschild black hole were derived up to 22PN~\cite{14PN,22PN} 
using a more systematic method to solve the Teukolsky equation~\cite{MST,MSTR}. 
It was shown that 
dephases between 22PN waveforms and very highly accurate waveforms 
during two-year inspirals can be less than $10^{-2}$ rad, 
and hence 22PN expressions might be used to detect gravitational waves 
from EMRIs. 
In this paper, by extending our previous results 
in Refs.~\cite{FI2010,PBFRT,14PN,22PN}, 
we derive the gravitational energy flux at 11PN for a test particle 
in circular orbits around the equatorial plane of a Kerr black hole 
and investigate how high PN order expressions 
for gravitational waves can improve the accuracy in PN results. 
We also obtain the gravitational energy flux into the horizon at 22.5PN 
for a test particle in circular orbits around a Schwarzschild black hole 
to fill the gap in the PN order between the energy flux at infinity, 
currently known at 22PN, and the horizon, previously known at 6.5PN 
beyond the Newtonian approximation~\cite{TMT}. 

The paper is organized as follows. 
In Sec.~\ref{sec:formulation}, we give a brief review of a formalism 
developed by Teukolsky and describe the necessary formulas in the paper. 
In Sec.~\ref{sec:PN_results}, we show analytic expressions for 
the energy flux to infinity at 7.5PN and to the horizon at 7PN 
beyond the Newtonian approximation. 
(The full analytic expressions will be shown online~\cite{BHPC}.) 
In order to investigate the accuracy of the PN results in the paper 
and the applicability to eLISA data analysis, 
comparisons between PN results and very accurate numerical results 
are done in Sec.~\ref{sec:comparison}. 
Section~\ref{sec:summary} is devoted to a summary and discussions. 
Finally in the appendices we describe supplemental equations 
to practically compute the formulas in Sec.~\ref{sec:formulation}. 
Throughout this paper, we use geometrized units with $c=G=1$. 
\section{Basic formulation}
\label{sec:formulation}
We solve the Teukolsky equation to calculate gravitational radiation from 
a particle orbiting around a Kerr black hole. In the Teukolsky formalism, 
the gravitational perturbation of the Kerr black hole is described by 
a master variable. If we consider the outgoing radiation to infinity, 
the master variable, the Newman-Penrose quantity $\Psi_4$, is related to 
the amplitude of the gravitational wave at infinity by
\begin{eqnarray}
\Psi_4\,\rightarrow\frac{1}{2}(\ddot{h}_{+}-i\,\ddot{h}_{\times })\,\,\,{\rm for}\,\,\,r\rightarrow\infty.
\end{eqnarray}

The Teukolsky equation can be solved by the decomposition of $\Psi_4$ as
\begin{eqnarray}
\Psi_4= {1\over (r-i a \cos\theta)^{4}}\,
\displaystyle \sum_{\ell,m}\int_{-\infty}^{\infty} d\omega\, 
e^{-i \omega t}\,\frac{e^{i m \varphi}}{\sqrt{2\pi}}\,{}_{-2}S_{\ell m}^{a\,\omega}(\theta)\,R_{\ell m\omega}(r),
\label{eq:psi4}
\end{eqnarray}
where $a$ is the angular momentum of 
the black hole and the angular function $_{-2}S_{\ell m}^{a\omega}(\theta)$ 
is the spin-weighted spheroidal harmonic with spin weight $s=-2$, 
normalized as 
\begin{eqnarray}
\int_0^{\pi} | \ _{-2}S_{\ell m}^{a\omega}(\theta)|^2 \sin\theta\,d\theta=1. 
\label{eq:normal_S}
\end{eqnarray}

The decomposition of $\Psi_4$ Eq.~(\ref{eq:psi4}) 
leads to the separation of the Teukolsky 
equation into radial and angular parts, 
\begin{align}
\Biggl[{1 \over \sin\theta}{d \over d\theta}
 \left\{\sin\theta {d \over d\theta} \right\}
-a^2\,\omega^2\,\sin^2\theta
 -{(m+s\,\cos\theta)^2 \over \sin^2\theta}~~~~~~~~~~~~~~~~~~~&\cr
-2\,a\,\omega s \cos\theta+s+2\,m\,a\,\omega+\lambda\Biggr] 
{}_{-2}S_{\ell m}^{a\omega}(\theta)&=0,\label{eq:Sph_costh}\\
\left[\Delta^2{d\over dr}\left({1\over \Delta}{d\over dr}\right)
-\left(-{K^2+4\,i\,(r-M)\,K \over \Delta}+8\,i\,\omega r+\lambda\right)\right]
 R_{\ell m\omega}(r)&=T_{\ell m\omega}(r),
\label{eq:Teu}
\end{align}
where $\lambda$ is the eigenvalue of $_{-2}S_{\ell m}^{a\omega}$, 
$\Delta=r^2-2Mr+a^2$, $K=(r^2+a^2)\omega-ma$ and $T_{\ell m\omega}$ is 
the source term of the particle. 

To solve Eq.~(\ref{eq:Teu}), we define two independent homogeneous 
solutions of the radial Teukolsky equation as 
\begin{align}
& R_{\ell m\omega}^{\rm in}=\left\{
  \begin{array}{lcc}
    B_{\ell m\omega}^{\rm trans}\,\Delta^2\,e^{-i k r^{*}} & \hbox{for} 
    & r^{*}\rightarrow -\infty, \\
    r^3\,B_{\ell m\omega}^{\rm ref}\,e^{i \omega r^{*}} +
    r^{-1}\,B_{\ell m\omega}^{\rm inc}\,e^{-i \omega r^{*}} & \hbox{for} 
    & r^{*}\rightarrow +\infty,
   \end{array}
\right. \cr
& R_{\ell m\omega}^{\rm up}=\left\{
\begin{array}{lcc}
C^{\rm up}_{\ell m\omega}\,e^{i k r^{*}}+\Delta^{2}\,C^{\rm ref}_{\ell m\omega}\,e^{-i k r^{*}}& \hbox{for} & r^{*}\rightarrow -\infty, \\
r^3\,C^{\rm trans}_{\ell m\omega}\,e^{i \omega r^{*}}&\hbox{for} & r^{*}\rightarrow +\infty,
\end{array}
\right. 
\label{eq:bc_rin_rup}
\end{align}
where $k=\omega - ma/(2Mr_+)$ and 
$r^*$ is the tortoise coordinate defined as
\begin{equation}
r^* = r + \frac{2Mr_+}{r_+-r_-}\ln\frac{r-r_+}{2M}-\frac{2Mr_-}{r_+-r_-}\ln\frac{r-r_-}{2M},
\end{equation}
with $r_{\pm}=M\pm\sqrt{M^2-a^2}$. 

Using the two independent solutions Eq.~(\ref{eq:bc_rin_rup}), 
with the Green function method 
one can construct a solution of the radial Teukolsky equation that 
is purely outgoing at infinity and purely incoming at the horizon 
\begin{align}
R_{\ell m \omega}(r)=
\frac{1}{W_{\ell m \omega}}\left\{
R^{\rm up}_{\ell m\omega}(r)
\int_{r_+}^r dr' {R^{\rm in}_{\ell m \omega}\,T_{\ell m \omega}\over\Delta^{2}}
+R^{\rm in}_{\ell m\omega}(r)
\int_{r}^\infty dr' {R^{\rm up}_{\ell m \omega}\,T_{\ell m \omega}\over\Delta^{2}}
\right\},
\label{eq:Rlmw}
\end{align}
where the Wronskian $W_{\ell m \omega}$ is given as
\begin{equation}
W_{\ell m \omega}=2\,i\,\omega\,C^{\rm trans}_{\ell m \omega}\,B^{\rm inc}_{\ell m \omega}.
\end{equation}

Then, the solution has the asymptotic form at the horizon as
\begin{eqnarray}
R_{\ell m \omega}(r\rightarrow r_+)=
\frac{B^{\rm trans}_{\ell m \omega}\,\Delta^2e^{-i k r^*}}
{2 i \omega C^{\rm trans}_{\ell m \omega}\,B^{\rm inc}_{\ell m \omega}}
\int_{r_+}^\infty dr' {R^{\rm up}_{\ell m \omega}\,T_{\ell m \omega}\over\Delta^{2}}
\equiv Z^{\rm H}_{\ell m \omega}\,\Delta^2\,e^{-i k r^*},
\label{eq:ZH}
\end{eqnarray}
while the solution has the following asymptotic form at infinity as
\begin{eqnarray}
R_{\ell m \omega}(r\rightarrow \infty)=
\frac{r^3\,e^{i\omega r^*}}{2 i \omega B^{\rm inc}_{\ell m \omega}}
\int_{r_+}^{\infty}dr'
{R^{\rm in}_{\ell m \omega}\,T_{\ell m \omega}\over\Delta^{2}}
\equiv Z^\infty_{\ell m \omega}\,r^3\,e^{i \omega r^*}.
\label{eq:Zinf}
\end{eqnarray}

Using the formula of the source term $T_{\ell m \omega}$~\cite{ST,chapter}, 
$Z_{\ell m\omega}^{\infty,{\rm H}}$ are expressed as
\begin{align}
Z^{\T{H}}_{\ell m\omega} &= 
\frac{\mu B^{\T{trans}}_{\ell m\omega}}{2i\omega C^{\T{trans}}_{\ell m\omega}B^{\T{inc}}_{\ell m\omega}}\int^{\infty}_{-\infty}dt\,e^{i \omega t-i m \phi(t)}\,\C{I}_{\ell m \omega}^{\T{H}}[r(t),\theta(t)],\cr
Z^{\infty}_{\ell m\omega} &= 
\frac{\mu}{2i\omega B^{\T{inc}}_{\ell m\omega}}\int^{\infty}_{-\infty}dt\,e^{i \omega t-i m \phi(t)}\,\C{I}_{\ell m \omega}^{\infty}[r(t),\theta(t)],
\label{eq:Z8H}
\end{align}
where 
\begin{align}
\C{I}^{\T{H}}_{\ell m\omega} =&
\left[ R^{\T{up}}_{\ell m\omega}\left\{A_{nn0}+A_{\bar{m}n0}+A_{\bar{m}\bar{m}0}\right\}
-\frac{dR^{\T{up}}_{\ell m\omega}}{dr}\left\{A_{\bar{m}n1}+A_{\bar{m}\bar{m}1}\right\}+\frac{d^2R^{\T{up}}_{\ell m\omega}}{d^2r}A_{\bar{m}\bar{m}2}\right]_{r=r(t),\theta=\theta(t)},\cr
 \C{I}^{\infty}_{\ell m\omega} =&
\left[ R^{\T{in}}_{\ell m\omega}\left\{A_{nn0}+A_{\bar{m}n0}+A_{\bar{m}\bar{m}0}\right\}
-\frac{dR^{\T{in}}_{\ell m\omega}}{dr}\left\{A_{\bar{m}n1}+A_{\bar{m}\bar{m}1}\right\}+\frac{d^2R^{\T{in}}_{\ell m\omega}}{d^2r}A_{\bar{m}\bar{m}2}\right]_{r=r(t),\theta=\theta(t)},
\label{eq:source}
\end{align}
and where $A_{nn0}$ and other terms are given in Appendix~\ref{sec:source}.

When the particle follows bound geodesics of Kerr spacetime, 
there exist three fundamental frequencies for the orbits~\cite{Schmidt:2002} 
and hence the frequency spectrum of $T_{\ell m\omega}$ becomes discrete.
In the case of circular orbits, $\tilde Z_{\ell m\omega}^{\infty,{\rm H}}$ 
in Eq.~(\ref{eq:Z8H}) takes the form 
\begin{equation}
 Z^{\infty,{\rm H}}_{\ell m\omega}=\tilde Z^{\infty,{\rm H}}_{\ell m\omega}\,\delta(\omega-m\,\Omega),
\label{eq:tildeZ}
\end{equation}
where $\Omega=v_r/(r_0(1+qv_r^3))$ is the angular frequency of the particle, 
$r_0$ is the orbital radius, $q=a/M$, and $v_r=\sqrt{M/r_0}$ is 
the orbital velocity. 

The time-averaged gravitational wave luminosity at infinity is 
then given by~\cite{TP1974} 
\begin{equation} 
 \left<{dE\over dt}\right>_\infty =\sum_{\ell=2}^{\infty} \sum_{m=-\ell}^{\ell}
   \frac{\vert \tilde Z^\infty_{\ell m\omega}\vert^2}{4\pi \omega^2}
   \equiv \left(dE\over dt\right)_{\rm N}\,\sum_{\ell=2}^{\infty}\,\sum_{m=-\ell}^{\ell}\,\eta_{\ell m}^\infty,
\label{eq:dEdt8}
\end{equation}
where $\left<\cdots\right>$ represents the time average, $\omega =m\Omega$, and 
$(dE/dt)_{\rm N}$ is the Newtonian quadrupole formula defined by 
\begin{equation}
\left(dE\over dt\right)_{\rm N}=\frac{32}{5}\left(\frac{\mu}{M}\right)^2 v^{10},
\label{eq:dEdtN}
\end{equation}
with $v\equiv (M\Omega)^{1/3}$. 
Similarly, the time-averaged gravitational wave luminosity 
at the horizon becomes~\cite{TP1974} 
\begin{equation}
 \left<{dE\over dt}\right>_{\rm H} =\sum_{\ell=2}^{\infty} \sum_{m=-\ell}^{\ell}
  \alpha_{\ell m\omega}\frac{\vert \tilde Z^{\rm H}_{\ell m\omega}\vert^2}{4\pi \omega^2}
  \equiv \left({dE\over dt}\right)_{\rm N}\,v^5\,\sum_{\ell=2}^{\infty}\,\sum_{m=-\ell}^{\ell}\,\eta_{\ell m}^{\rm H}, 
\label{eq:dEdtH}
\end{equation}
where 
\begin{equation}
 \alpha_{\ell m\omega} =
 \frac{256(2Mr_+)^5 k (k^2+4\tilde\epsilon^2)(k^2+16\tilde\epsilon^2)\omega^3}
{|C|^2},
\label{eq:alphaH}
\end{equation}
with $\tilde\epsilon=\sqrt{M^2-a^2}/(4Mr_{+})$ and 
\begin{eqnarray*}
|C|^2 =& \left[(\lambda+2)^2+4\,a\,\omega\,m-4\,a^2\,\omega^2\right]\,
       \left[\lambda^2+36\,a\,\omega\,m-36\,a^2\,\omega^2\right]\,\cr
       & +(2\,\lambda+3)\,(96\,a^2\,\omega^2-48\,a\,\omega\,m)
        +144\,\omega^2\,(M^2-a^2).
\end{eqnarray*}

Finally, the gravitational waveforms are given in terms of 
$\tilde Z_{\ell m\omega}^{\infty}$ as 
\begin{eqnarray}
h_{+}-i\,h_{\times }=-\frac{2}{r}\,\sum _{\ell,m} \frac{\tilde Z^\infty_{\ell m\omega}}{\omega^2}\frac{e^{i m \varphi}}{\sqrt{2\pi}}\,_{-2}S_{\ell m}^{a\omega}(\theta)\,e^{i \omega (r^{*}-t)}.
\label{eq:hpm_slm}
\end{eqnarray}

In this paper, 
using Eqs.~(\ref{eq:dEdt8}), (\ref{eq:dEdtH}) and (\ref{eq:hpm_slm}) 
we compute the gravitational energy flux and waveforms 
in the post-Newtonian approximation, i.e., in the expansion 
with respect to $v= (M\Omega)^{1/3}$. 
For this purpose, it is necessary to compute the asymptotic amplitudes 
$\tilde Z_{\ell m\omega}^{\infty,{\rm H}}$, which involve calculations of 
the angular Teukolsky function $_{-2}S_{\ell m}^{a\omega}(\theta)$ and 
the radial Teukolsky functions $R^{\rm in,up}_{\ell m\omega}(r)$. 
To this end, in Appendices~\ref{sec:PN_Sph} and \ref{sec:PN_MST}, 
we give a short review of the calculations of the series expansions of
$_{-2}S_{\ell m}^{a\omega}(\theta)$ and $R^{\rm in,up}_{\ell m\omega}(r)$ 
in terms of $\epsilon\equiv 2M\omega=O(v^3)$ and $z=\omega r=O(v)$. 
\section{11PN results for the time-averaged energy flux}
\label{sec:PN_results}
In this paper, we derive the 11PN formula for the energy flux in the case 
of a test particle in a circular orbit around the equatorial plane of 
a Kerr black hole. Since the expressions are very long, 
we exhibit the 7.5PN expression for the energy flux at infinity 
in Sec.~\ref{sec:dEdt8} and the next new 7PN terms in the energy flux into 
the horizon in Sec.~\ref{sec:dEdtH}. 
We also compute the energy flux into the event horizon for a particle 
in a circular orbit around a Schwarzschild black hole up to 
22.5PN beyond the Newtonian approximation. 
The complete expressions for the energy flux 
will be publicly available online~\cite{BHPC}. 
\subsection{Infinity flux}
\label{sec:dEdt8}
The 7.5PN energy flux to infinity is given by
\begin{eqnarray}
  \left\langle{dE \over dt}\right\rangle_\infty &=&
  \left( {dE \over dt} \right)_\mathrm{N}
  \Biggl[ 1 + \left\{ q\mbox{-independent terms} \right\} 
  - {11 \over 4} q v^3 + {33 \over 16} q^2 v^4
  - {59 \over 16} q v^5
  \cr
  && + \left\{ - {65 \over 6} \pi q +
  {611 \over 504} q^2 \right\} v^6 +
  \left\{ {162035 \over 3888} q + {65 \over 8} \pi q^2 -
  {71 \over 24} q^3 \right\} v^7
  \cr
  && + \left\{ - {359 \over 14} \pi q + {22667 \over 4536} q^2 +
  {17 \over 16} q^4 \right\} v^8 
  \cr
  && + \biggl\{
  \left(- { \frac {9828207709}{52390800}} + { \frac {40939}{315}} \,\ln 2 - { \frac {43}{3}} \,\pi^{2} + { \frac {6841}{105}} \,\gamma   + { \frac {6841}{105}} \,\ln v \right)\,q  
  \cr
  &&\qquad\;\,
  + { \frac {8447}{672}} \,\pi \,q^{2} - { \frac {112025}{4536}} \,q^{3}\biggr\} v^9
  \cr
  && + \biggl\{    \frac {23605}{144} \,\pi \,q  
+ \left({ \frac {93301799461}{628689600}}  - { \frac {27499}{420}} \,\ln 2 + { \frac {43}{4}} \,\pi^{2} - { \frac {4601}{140}} \,\gamma 
- { \frac {4601}{140}} \,\ln v\right)\,q^{2} 
  \cr
  &&\qquad\;\,
  - { \frac {45}{4}} \,\pi \,q^{3} + { \frac {731}{126}} \,q^{4}
\biggr\} v^{10}
  \cr
  && + \biggl\{
\biggl( - { \frac {244521688471}{272432160}} - { \frac {1280791}{10584}} \,\ln 2 - { \frac {671}{12}} \,\pi^{2} + { 
\frac {128459}{7560}} \,\gamma  + { \frac {486243}{
3136}} \,\ln 3
  \cr
  &&\qquad\;\,
  + { \frac {128459}{7560}} \,\ln v\biggr)q
 + { \frac {34211}{1512}} \,\pi \,q^{2} - 
{ \frac {257407}{9072}} \,q^{3} + { 
\frac {33}{8}} \,\pi \,q^{4}  - { \frac {1}{8}} \,q^{5}
\biggr\} v^{11}
  \cr
  && + \biggl\{
 \left( - { \frac {270159823411}{558835200}} \,
\pi  + { \frac {81878}{315}} \,\pi \,\gamma  + 
{ \frac {54514}{105}} \,\pi \,\ln 2 + 
{ \frac {81878}{315}} \,\pi \,\ln v\right)\,q 
  \cr
  &&\qquad\;\,
+ \biggl(
{ \frac {13501670684927}{28605376800}}  - 
{ \frac {22901}{196}} \,\gamma  + { 
\frac {24763}{756}} \,\pi^{2} - { \frac {537013}{
3780}} \,\ln 2 
  \cr
  &&\qquad\;\,
- { \frac {142155}{1568}} \,\ln 3 
 - { \frac {22901}{196}} \,\ln v\biggr)q^{2} 
- { \frac {67426}{567}} \,\pi \,q^{3} + 
{ \frac {24397}{1008}} \,q^{4} 
\biggr\} v^{12}
\cr&&\cr&&\vspace{6pt}
+ \biggl\{
\left( {\frac {256}{15}}\,\ln \kappa -{\frac {
5160697541}{6735960}}\,\gamma-{\frac {16342453091}{47151720}}\,\ln 2 
-{\frac {67221333}{86240}}\,\ln 3 
\right.
\cr&&\cr&&\vspace{6pt}\hspace{0.6cm}
\left.
+{\frac {526805}{2916}}\,{\pi }^{2}
+{\frac {1290587071610633}{606842636400}} \right) q 
\cr&&\cr&&\vspace{6pt}\hspace{0.6cm}
+ \left( -{\frac {18297}{70}}\,\pi \,\ln 2 
-{\frac {27499}{210}}\,\pi \,\gamma+{\frac {
1241484285313}{2235340800}}\,\pi  \right) {q}^{2}
\cr&&\cr&&\vspace{6pt}\hspace{0.6cm}
+ \left( -{\frac {
416537225257}{1414551600}}-{\frac {263}{18}}\,{\pi }^{2}+{\frac {31467
}{140}}\,\ln 2 +{\frac {256}{5}}\,\ln \kappa 
+{\frac {174659}{1260}}\,\gamma \right) {q}^{3}
\cr&&\cr&&\vspace{6pt}\hspace{0.6cm}
+{\frac {60869}{2016}}\,\pi \,{q}^{4}
+{\frac {11311}{1008}}\,{q}^{5}
+ \left( {\frac {256}{15}}\,q+{\frac {256}{5}}\,{q}^{3} \right) \Psi_{\rm A}^{(0,2)}\left(q \right) 
\cr&&\cr&&\vspace{6pt}\hspace{0.6cm}
+ \left( 
-{\frac {5045737157}{6735960}}\,q-{\frac {27499}{210}}\,\pi \,{q}^{2}+{\frac {239171}{1260}}\,{q}^{3} \right) \ln v
\biggr\} v^{13}
\cr&&\cr&&\vspace{6pt}
+ \biggl\{
\left( -{\frac {2628337}{15120}}\,
\pi \,\ln  2 +{\frac {1458729}{1568}}\,\pi \,\ln 3 
+{\frac {8005397}{21168}}\,\pi \,\gamma-{\frac {748453609847597}{152562009600}}\,\pi  \right) q
\cr&&\cr&&\vspace{6pt}\hspace{0.6cm}
+ \left( -{\frac {921209351}{7858620}}\,\gamma-{\frac {5442958727}{7858620}}\,\ln 2 +{\frac {206753}{3402}}\,{\pi }^{2}+{\frac {6830001}
{34496}}\,\ln 3 
\right.
\cr&&\cr&&\vspace{6pt}\hspace{0.6cm}\hspace{0.6cm}
\left.
+{\frac {3708535586671457}{2831932303200}} \right) {q}^{2}
-{\frac {238825}{1008}}\,\pi \,{q}^{3}
\cr&&\cr&&\vspace{6pt}\hspace{0.6cm}
+ \left( -{\frac {1391}{84}}\,\gamma-{
\frac {4601}{140}}\,\ln 2 +{\frac {65}{12}}\,{\pi }^{2
}+{\frac {305698147}{1270080}} \right) {q}^{4}
-\frac{1}{4}\,\pi \,{q}^{5}+{\frac {827}{336}}\,{q}^{6}
\cr&&\cr&&\vspace{6pt}\hspace{0.6cm}
+ \left( 
{\frac {8005397}{21168}}\,\pi \,q
-{\frac {921209351}{7858620}}\,{q}^{2}-{\frac {1391}{84}}\,{q}^{4} \right) \ln v
\biggr\} v^{14}
\cr&&\cr&&\vspace{6pt}
+ \biggl\{
\left( -{\frac {18591892}{6615}}\,\gamma\,\ln 2 +{
\frac {1316293890558029}{235994358600}}\,\ln 2 +{
\frac {1154100056435789}{235994358600}}\,\gamma
\right.
\cr&&\cr&&\vspace{6pt}\hspace{0.6cm}\hspace{0.6cm}
+{\frac {109028}{315}}\,{\pi }^{2}\gamma
+{\frac {76}{5}}\,{\pi }^{4}+{\frac {1195923689}{
15717240}}\,{\pi }^{2}-{\frac {7750438}{11025}}\,{\gamma}^{2}-{\frac {
30978854}{11025}}\, \left( \ln 2 \right)^{2}
\cr&&\cr&&\vspace{6pt}\hspace{0.6cm}\hspace{0.6cm}
-{\frac {632}{35}}\,\ln \kappa +{\frac {6197998046875}{5479778304}}\,\ln 5 
+{\frac {130748}{189}}\,{\pi }^{2}
\ln 2 -{\frac {36461702637}{157853696}}\,\ln 3 
\cr&&\cr&&\vspace{6pt}\hspace{0.6cm}\hspace{0.6cm}
\left.
+{\frac {109028}{105}}\,\zeta  \left( 3 \right) -{\frac {
2049619936309873009541}{1807162200133290000}} \right) q
\cr&&\cr&&\vspace{6pt}\hspace{0.6cm}
+ \left( {
\frac {418783053049409}{152562009600}}\,\pi -{\frac {426465}{784}}\,
\pi \,\ln 3 -{\frac {13188979}{17640}}\,\pi \,\ln 2 
-{\frac {34227289}{52920}}\,\pi \,\gamma \right) {q}^{2}
\cr&&\cr&&\vspace{6pt}\hspace{0.6cm}
+ \left( {\frac {1079649}{3136}}\,\ln 3 +{\frac {
166255711}{238140}}\,\ln 2 
-{\frac {681171487193473}{257448391200}}
-{\frac {683477}{3402}}\,{\pi }^{2}
\right.
\cr&&\cr&&\vspace{6pt}\hspace{0.6cm}\hspace{0.6cm}
\left.
+{\frac {118154591}{238140}}\,\gamma
-{\frac {5618}{105}}\,\ln \kappa  \right) {q}^{3}
+{\frac {1790339}{13608}}\,\pi \,{q}^{4}
-{\frac {10080445}{81648}}\,{q}^{5}
\cr&&\cr&&\vspace{6pt}\hspace{0.6cm}
+ \left( -{\frac {8}{45}}\,q+\frac{2}{15}\,{q}^{3} \right) \Psi_{\rm A}^{(0,1)}(q) 
+ \left( -{\frac {5632}{315}}\,q-{\frac {5632}{105}}\,{q}^{3} \right) \Psi_{\rm A}^{(0,2)}(q) 
\cr&&\cr&&\vspace{6pt}\hspace{0.6cm}
+ 
\left(\left( {\frac {109028}{315}}\,{\pi }^{2}+{
\frac {1149838672589069}{235994358600}}-{\frac {18591892}{6615}}\,\ln 2 
-{\frac {15500876}{11025}}\,\gamma \right) q
\right.
\cr&&\cr&&\vspace{6pt}\hspace{0.6cm}\hspace{0.6cm}
\left.\left.
-{\frac {34227289}{52920}}\,\pi \,{q}^{2}
+{\frac {105412967}{238140}}\,{q}^{3}
 \right)\right) \ln v
-{\frac {7750438}{11025}}\,q 
(\ln v)^{2}
\biggr\} v^{15}
 \Biggr],
\label{eq:dEdt8_x15}
\end{eqnarray}
where $\kappa=\sqrt{1-q^2}$, $\gamma$ is the Euler constant, 
$\zeta(n)$ is the zeta function, 
\begin{align*}
\Psi_{\rm A}^{(n,m)}(q)&=\frac{1}{2}\left[\Psi^{(n)}\left(3+\frac{i\,m\,q}{\sqrt{1-q^2}}\right)+\Psi^{(n)}\left(3-\frac{i\,m\,q}{\sqrt{1-q^2}}\right)\right],
\cr
\Psi_{\rm B}^{(n,m)}(q)&=\frac{1}{2\,i}\left[\Psi^{(n)}\left(3+\frac{i\,m\,q}{\sqrt{1-q^2}}\right)-\Psi^{(n)}\left(3-\frac{i\,m\,q}{\sqrt{1-q^2}}\right)\right],
\end{align*}
and $\Psi^{(n)}(z)$ is the polygamma function. 

The ${\mathcal O}(v^{9})-{\mathcal O}(v^{15})$ terms in Eq.~(\ref{eq:dEdt8_x15}) 
are the new terms derived by the post-Newtonian approximation in this paper.
\footnote{We note that the ``$q$-independent terms'' 
in Eq.~(\ref{eq:dEdt8_x15}) coincide with those 
for the Schwarzschild black hole, since the $q$-dependent terms in 
Eq.~(\ref{eq:dEdt8_x15}), 
{\it e.g.} $v^j\,q^k\,\ln\kappa$ and $v^j\,q^k\,\Psi_{\rm A,B}^{(n,m)}(q)$ 
where $j$ and $k$ are integers, vanish when $q=0$. 
For 8PN and higher PN orders, however, 
there are $q$-dependent terms, 
{\it e.g.} $v^j\,\kappa$ and $v^j\,\Psi_{\rm A}^{(n,m)}(q)$, 
which do not vanish when $q=0$ and hence the ``$q$-independent terms''
do not agree with those for the Schwarzschild black hole.} 
Among these terms, the ${\mathcal O}(v^{9})-{\mathcal O}(v^{11})$ terms 
agree with the analytic expressions in Ref.~\cite{Shah_20PN}, 
which determined the post-Newtonian coefficients of the energy flux up to 20PN 
by fitting with very accurate, one part in $10^{600}$, 
numerical calculation of the energy flux. 
For the 6PN and higher PN order energy flux at infinity, 
in Ref.~\cite{Shah_20PN}, 
some of the post-Newtonian coefficients are not extracted as analytic values 
but as numerical values. 
This is not only because it is difficult to numerically extract 
analytic coefficients for combinations of transcendental numbers such as 
$\pi$, Euler's constant, and logarithms of prime numbers, 
but also because numerical fitting of post-Newtonian coefficients is done 
by presenting these coefficients as a polynomial in $q$ 
although irrational functions in $q$ such as polygamma functions and 
logarithms appear from 6.5PN onward as shown in Eq.~(\ref{eq:dEdt8_x15}). 
Further, by performing a small $q$ expansion of our 11PN expression, 
we also find that our 11PN energy flux to infinity is consistent with the one 
in Ref.~\cite{Shah_20PN} up to 11PN. 
\footnote{
It might be noted from Eq.~(\ref{eq:dEdt8_x15}) that, if one includes 
$\kappa$, $\ln\kappa$, and $\Psi_{\rm A,B}^{(n,m)}(q)$ for the numerical fitting 
in Ref.~\cite{Shah_20PN}, 
one might be able to obtain a more accurate fitting formula.}

From Eq.~(\ref{eq:dEdt8_x15}), we find the coefficient in $q\,(\ln v)^0$ 
at 6PN is given by 
\begin{equation}
- { \frac {270159823411}{558835200}} \,
\pi  + { \frac {81878}{315}} \,\pi \,\gamma  + 
{ \frac {54514}{105}} \,\pi \,\ln 2. 
\end{equation}
The above analytic value is consistent with the numerical value of 
the coefficient in $q\,(\ln v)^0$ at 6PN energy flux to infinity 
in Ref.~\cite{Shah_20PN}, $83.16039023577041\ldots$.
\subsection{Horizon flux}
\label{sec:dEdtH}
The next new 7PN terms for the energy flux into the horizon are given by 
\begin{eqnarray}
  \left\langle{dE \over dt}\right\rangle_{\rm H}^{(9)}&=&
{\frac {2204129}{22050}}-{4\,{\pi }^{2}\over 3}
-{\frac {856}{105}}\,\gamma-{\frac {856}{105}}\,\ln 2 
-{\frac {856}{105}}\,\ln \kappa 
\cr&&\cr&&\vspace{6pt}
-{\frac {856}{105}}\,{\frac {\ln \kappa }{\kappa}}-{\frac {856}{105}}\,{\frac {\ln 2 }{\kappa}}-{\frac {856}{105}}\,{\frac {\gamma}{\kappa}}-{\frac {4\,{\pi }^{2}}{3\kappa}}+{\frac {2204129}{22050\kappa}}
\cr&&\cr&&\vspace{6pt}
+ \left\{ -{\frac {3424}{35}}\,{\frac {\ln 2}{\kappa}}+{\frac {24687352}{33075}}-{\frac {3424}{35}}\,{\frac {\ln \kappa}{\kappa}}
-{\frac {3424}{35}}\,{\frac {\gamma}{\kappa}}
-{\frac {30356}{315}}\,\ln 2 -16\,{\frac {{\pi }^{2}}{\kappa}}
\right.\cr&&\cr&&\vspace{6pt}\hspace{0.6cm}\left.
+{\frac {7205413}{9800\kappa}}
-{\frac {30356}{315}}\,\gamma
-{\frac {134}{9}}\,{\pi }^{2}-{\frac {30356}{315}}\,\ln \kappa \right\} {q}^{2}
\cr&&\cr&&\vspace{6pt}
+ \left\{ {\frac {856}{15}}\,{\frac {\ln 2 }{\kappa}}-{\frac {50225669}{176400}}+{\frac {28}{3}}\,{\frac {{\pi }^{2}}{\kappa}}-{\frac {1573}{105}}\,\gamma+{
\frac {856}{15}}\,{\frac {\gamma}{\kappa}}+{{\pi }^{2}\over 6}
+{\frac {856}{15}}\,{\frac {\ln \kappa }{\kappa}}-{\frac {2315279}{3150\kappa}}
\right.\cr&&\cr&&\vspace{6pt}\hspace{0.6cm}\left.
-{\frac {1573}{105}}\,\ln 2 -{\frac {1573}{105}}\,\ln \kappa \right\} {q}^{4}
\cr&&\cr&&\vspace{6pt}
+ \left\{ {\frac {19195}{168}}+{\frac {1712}{35}}\,{\frac {\ln \kappa }{\kappa}}+{\frac {8\,{\pi }^{2}}{\kappa}}-{\frac {2293757}{29400\kappa}}
+{\frac {1712}{35}}\,{\frac {\gamma}{\kappa}}+{\frac {1712}{35}}\,{\frac {\ln 2}{\kappa}} \right\} {q}^{6}-{\frac {621}{28}}\,{\frac {{q}^{8}}{\kappa}}
\cr&&\cr&&\vspace{6pt}
+ \left\{ -{\frac {428}{315}}\,{q}^{2}+{\frac {107}{105}}\,{q}^{4} \right\} {\Psi_{\rm A}^{(0,1)}}(q) 
\cr&&\cr&&\vspace{6pt}
+ \left\{ -{\frac {856}{105}}-{\frac {856}{105\kappa}}
+ \left( -{\frac {9976}{105}}-{\frac {3424}{35\kappa}} \right) {q}^{2}
+ \left( -16+{\frac {856}{15\kappa}} \right) {q}^{4}
+{\frac {1712}{35}}\,{\frac {{q}^{6}}{\kappa}}
\right.\cr&&\cr&&\vspace{6pt}\hspace{0.6cm}\left.
+ \left(-{\frac {3424}{105}}\,q-{\frac {3424}{35}}\,{q}^{3} \right) {\Psi_{\rm B}^{(0,2)}}(q)\right\} {\Psi_{\rm A}^{(0,2)}}
\cr&&\cr&&\vspace{6pt}
+ \left\{
  \left( -{16\over {\kappa}^{3}}-{\frac {64}{3{\kappa}^{2}}} \right) q
+ \left(-{32\over {\kappa}^{3}}-{\frac {176}{3{\kappa}^{2}}} \right) {q}^{3}
+ \left( {48\over {\kappa}^{3}}+{16\over {\kappa}^{2}} \right) {q}^{5} 
  \right\} {\Psi_{\rm B}^{(2,2)}}(q) 
\cr&&\cr&&\vspace{6pt}
+ \left\{
  \left( {4\over 3}+{8\over 3\kappa} \right) {q}^{2}
+ \left( -1-{2\over {\kappa}} \right) {q}^{4} 
  \right\} {\Psi_{\rm A}^{(1,1)}}(q) 
\cr&&\cr&&\vspace{6pt}
+ \left\{ 
  \left( -{\frac {3424}{105}}\,\ln \kappa +{224\over {\kappa}}-{\frac {3424}{105}}\,\ln 2 -{16\,{\pi }^{2}\over 3}+{\frac {4599008}{11025}}-{\frac {3424}{105}}\,\gamma \right) q
\right.\cr&&\cr&&\vspace{6pt}\hspace{0.6cm}
+ \left( -{\frac {3424}{35}}\,\ln 2 -16\,{\pi }^{2}-{\frac {3424}{35}}\,\ln \kappa -{\frac {3424}{35}}\,\gamma+{\frac {815386}{1225}}-{32\over {\kappa}} \right) {q}^{3}
\cr&&\cr&&\vspace{6pt}\hspace{0.6cm}\left.
+ \left( -{\frac {1774}{7}}-{288\over {\kappa}} \right) {q}^{5}
+96\,{\frac {{q}^{7}}{\kappa}} 
  \right\} {\Psi_{\rm B}^{(0,2)}}(q) 
\cr&&\cr&&\vspace{6pt}
+ \left\{ {16\over {\kappa}}+16+ \left( {192\over {\kappa}}+176 \right) {q}^{2}-112\,{\frac {{q}^{4}}{\kappa}}-96\,{\frac {{q}^{6}}{\kappa}} \right\}  \left\{ {\Psi_{\rm B}}^{(0,2)}(q) \right\}^{2}
\cr&&\cr&&\vspace{6pt}
+ \left\{ {\frac {135}{28}}\,q+{\frac {4995}{112}}\,{q}^{3}+{\frac {675}{14}}\,{q}^{5} \right\} {\Psi_{\rm B}^{(0,3)}}(q) + \left\{ {\frac {1712}{105}}\,{\frac {q}{\kappa}}+{\frac {1712}{35}}\,{\frac {{q}^{3}}{\kappa}} \right\} {\Psi_{\rm B}^{(1,2)}}(q) 
\cr&&\cr&&\vspace{6pt}
+ \left\{  \left( {\frac {6151}{252}}+{8\over 3\,{\kappa}} \right) q+ 
\left( {2\over {\kappa}}-{\frac {1437}{112}} \right) {q}^{3}
+ \left( -{\frac {26}{3\kappa}}-{\frac {87}{28}} \right) {q}^{5}
+4\,{\frac {{q}^{7}}{\kappa}} \right\} {\Psi_{\rm B}^{(0,1)}}(q) 
\cr&&\cr&&\vspace{6pt}
+ \left\{ {8\over 3}\,{q}^{2}-2\,{q}^{4} \right\}  \left\{ {\Psi_{\rm B}^{(0,1)}}(q)\right\}^{2}
\cr&&\cr&&\vspace{6pt}
+ \left\{ {24\over {\kappa}}+24+ \left( 296+{272\over {\kappa}} \right) {q}^{2}+ \left( -{56\over {\kappa}}+96 \right) {q}^{4}-48\,{\frac {{q}^{6}}{\kappa}}
\right.\cr&&\cr&&\vspace{6pt}\hspace{0.6cm}\left.
+ \left(  \left( 32+{64\over {\kappa}} \right) q+ 
\left( 96+{192\over {\kappa}} \right) {q}^{3} \right) {\Psi_{\rm B}^{(0,2)}}(q)  \right\} {\Psi_{\rm A}^{(1,2)}}(q) 
\cr&&\cr&&\vspace{6pt}
+ \left\{ {\frac {64}{3}}\,q+64\,{q}^{3} \right\}  \left\{ {\Psi_{\rm B}^{(0,2)}}(q) \right\}^{3}
\cr&&\cr&&\vspace{6pt}
+ \left\{ -{\frac {1712}{105\kappa}}-{\frac {1712}{105}}
+ \left( -{\frac {6848}{35\kappa}}-{\frac {60712}{315}} \right) {q}^{2}
+ \left( {\frac {1712}{15\kappa}}-{\frac {3146}{105}} \right) {q}^{4}
+{\frac {3424}{35}}\,{\frac {{q}^{6}}{\kappa}}
\right.\cr&&\cr&&\vspace{6pt}\hspace{0.6cm}\left.
+ \left( -{\frac {6848}{105}}\,q-{\frac {6848}{35}}\,{q}^{3} \right) {\Psi_{\rm B}^{(0,2)}}(q)  \right\} \ln v,
\label{eq:dEdtH_x9}
\end{eqnarray}
where $\left\langle{dE \over dt}\right\rangle_{\rm H}^{(9)}$ is defined through 
\begin{equation*}
\left\langle{dE \over dt}\right\rangle_{\rm H}=
\left(dE\over dt\right)_{\rm N}\,v^5\,\sum_{n=0}^{2 N_{\rm PN}-5} \, 
\left\langle{dE \over dt}\right\rangle_{\rm H}^{(n)}\,v^n, 
\end{equation*}
and $N_{\rm PN}$ is the PN order, i.e. $N_{\rm PN}=11$ when the PN order is 11PN. 
Notice that the energy flux down into the horizon starts at 
${\mathcal O}(v^5)$ (${\mathcal O}(v^8)$), i.e. 2.5PN (4PN), 
beyond the quadrupole formula when $q\neq 0$ ($q=0$)~\cite{TMT,PS1995}.

Again, performing a small $q$ expansion of our 11PN expression, 
we find our 11PN energy flux to the horizon is consistent with the one 
in Ref.~\cite{Shah_20PN} up to 11PN. 
For the case of a particle in a circular orbit around 
a Schwarzschild black hole, we also derive the energy flux 
down the event horizon at 22.5PN, which is consistent with the one 
in Ref.~\cite{Shah_20PN} up to 22.5PN. 
\section{Comparisons between 11PN results and numerical results}
\label{sec:comparison}
To investigate the accuracy of the energy flux in the 
post-Newtonian approximation, 
we compare PN results with numerical results, 
based on a method in Refs.~\cite{FT1,FT2}. With this numerical method, 
one can investigate gravitational waves with an accuracy of about 
$14$ significant figures in double precision calculations. 
Hence one can use the numerical results to estimate the accuracy in 
the PN results by a comparison. For the comparison in this section, 
we need to compute the energy flux using Eqs.~(\ref{eq:dEdt8}) 
and (\ref{eq:dEdtH}). 
To numerically compute the energy flux, we set the maximum value of 
$\ell$ to $15$, which gives the relative error in the energy flux 
better than $10^{-5}$ for the comparisons in this section. 
For the energy flux at 11PN, 
we need to compute $\ell$ up to 13 (5) for the energy flux to infinity 
(the horizon). 

In Sec.~\ref{sec:Taylor}, comparisons for the the energy flux are done 
for several values of the spin of the Kerr black hole. 
In Secs.~\ref{sec:Factorized} and \ref{sec:Exponential}, 
the same comparisons are done using resummation techniques, 
the factorized resummation introduced in Ref.~\cite{DIN} and 
the exponential resummation in Ref.~\cite{Isoyama:2012bx}, 
for the post-Newtonian approximation to the energy flux. 
We will see how resummation methods improve the performance 
in the post-Newtonian approximation for the energy flux. 
Finally, in Sec~\ref{sec:dephase}, 
we compare the total cycle of orbits during a two-year inspiral 
for representative binaries in the eLISA frequency band. 
\subsection{Energy flux: Taylor expanded PN approximation}
\label{sec:Taylor}
\begin{figure}[t]
\begin{center}
\includegraphics[width=69mm]{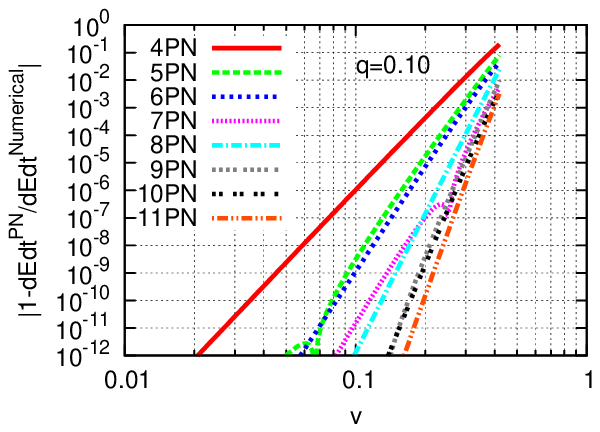}%
\includegraphics[width=69mm]{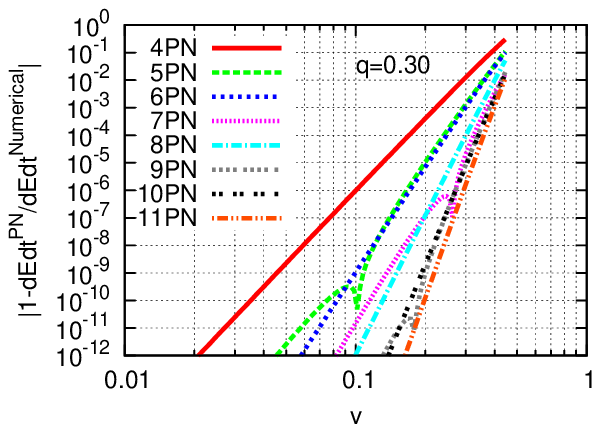}\\
\includegraphics[width=69mm]{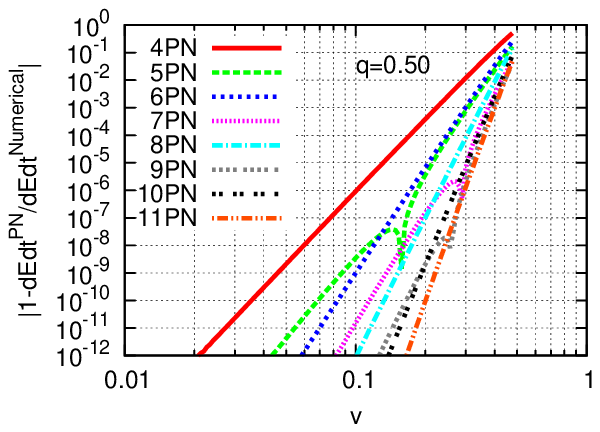}%
\includegraphics[width=69mm]{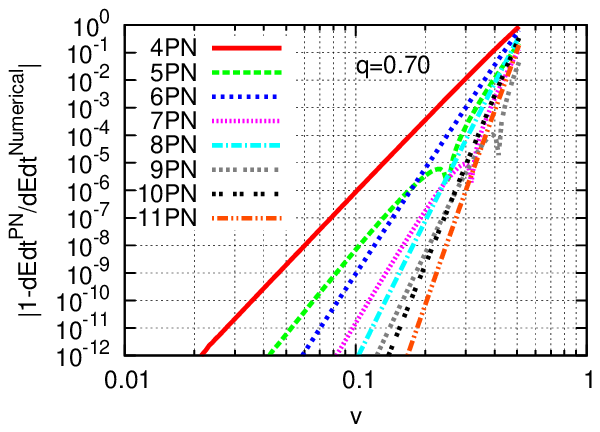}\\
\includegraphics[width=69mm]{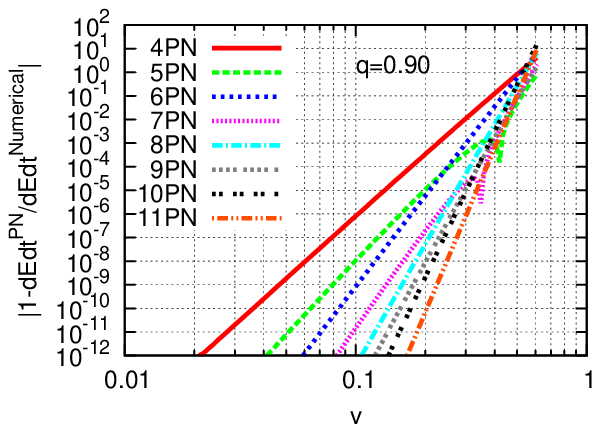}
\includegraphics[width=69mm]{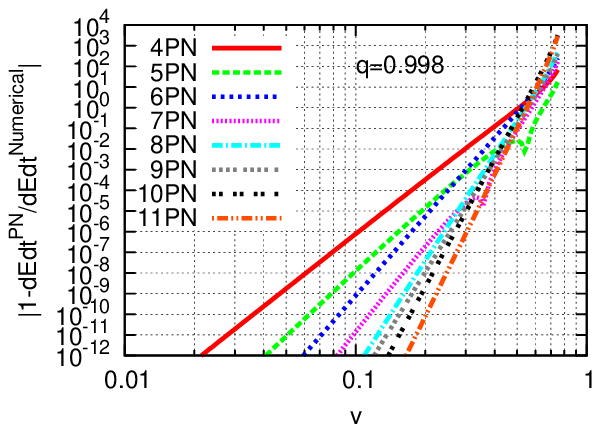}%
\end{center}
\caption{Absolute values of the relative error in the total energy flux 
from numerical results and PN approximations 
as a function of the orbital velocity, 
$v=(M/r_0)^{1/2}[1+q\,(M/r_0)^{3/2}]^{-1/3}$, 
up to ISCO for $q=0.1,\,0.3,\,0.5,\,0.7,\,0.9$ and $0.998$. 
The relative error for 11PN is smaller than $10^{-5}$ when $v\lessapprox 0.33$. 
}\label{fig:flux_taylor_1}
\end{figure}

\begin{figure}[t]
\begin{center}
\includegraphics[width=69mm]{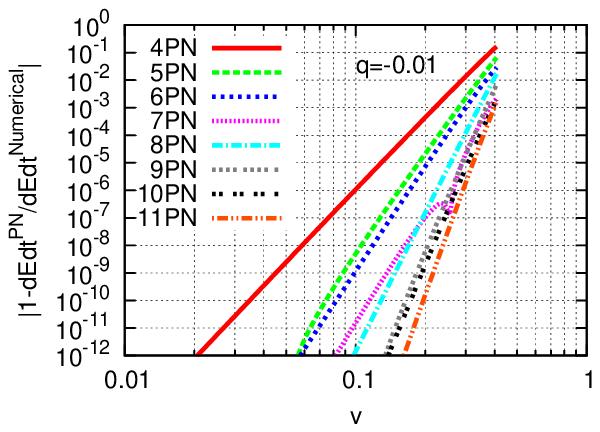}%
\includegraphics[width=69mm]{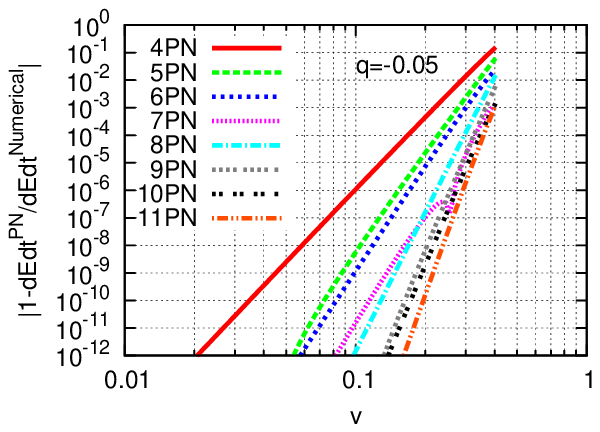}\\
\includegraphics[width=69mm]{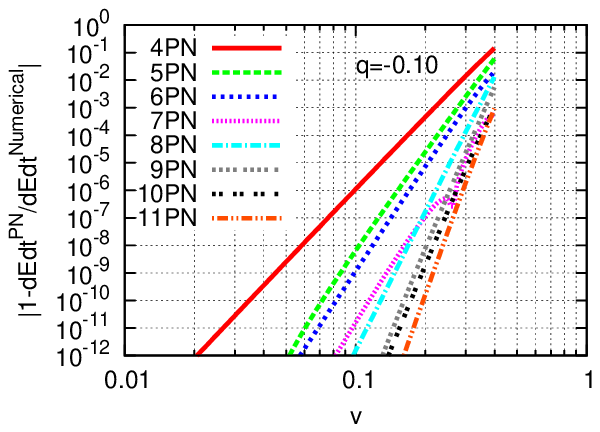}%
\includegraphics[width=69mm]{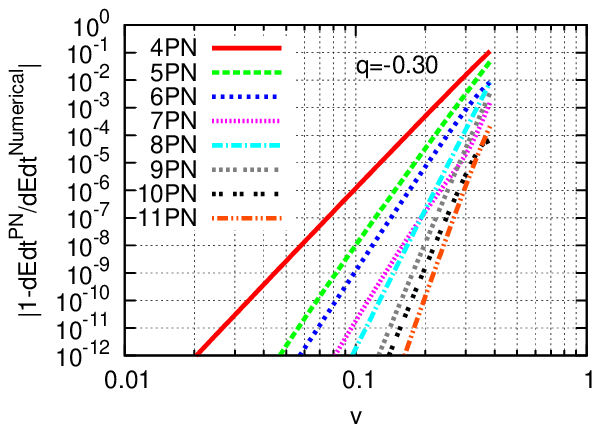}\\
\includegraphics[width=69mm]{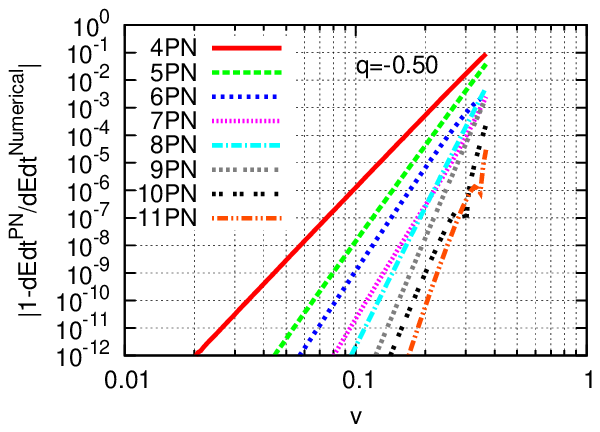}%
\includegraphics[width=69mm]{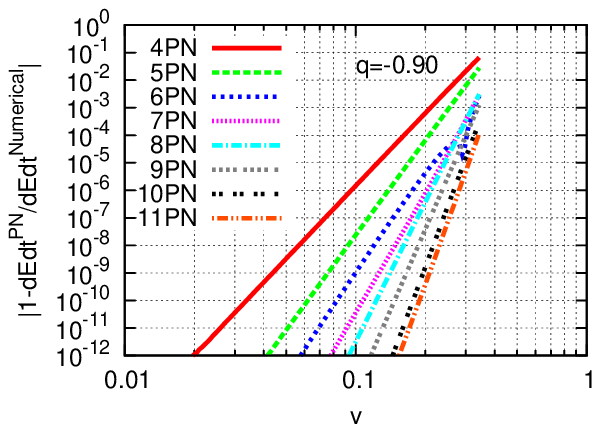}
\end{center}
\caption{Same as Fig.~\ref{fig:flux_taylor_1} but for 
$q=-0.01,\,-0.05,\,-0.1,\,-0.3,\,-0.5$ and $-0.9$. 
The relative error for 11PN is smaller than $10^{-5}$ when $v\lessapprox 0.33$. 
}\label{fig:flux_taylor_2}
\end{figure}

Figures~\ref{fig:flux_taylor_1} and \ref{fig:flux_taylor_2} show 
the relative error in the total energy flux from numerical results and 
PN approximations as a function of the orbital velocity up to 
the innermost stable circular orbit (ISCO). 
From these figures, one will find that 
the relative error becomes smaller with increasing PN order 
for $v\le 0.3$, 
except for accidental agreements for certain values of the velocity. 
However, the relative error around ISCO does not necessarily become smaller 
with increasing PN order when $q > 0.3$. 
The relative error for 11PN is smaller than $10^{-5}$ when $v\lessapprox 0.33$, 
irrespective of the values of $q$ investigated in the paper. 

\begin{figure}[h]
\begin{center}
\includegraphics[width=69mm]{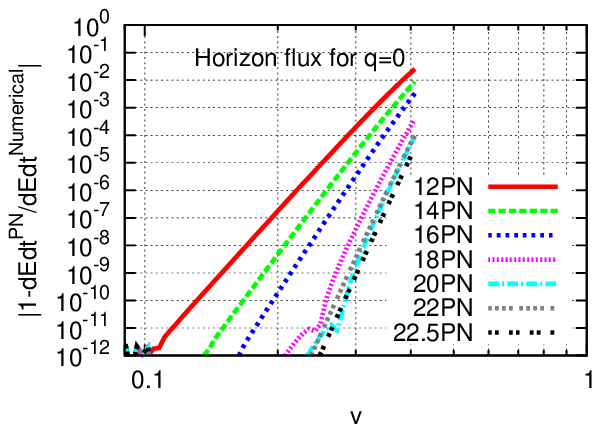}%
\includegraphics[width=69mm]{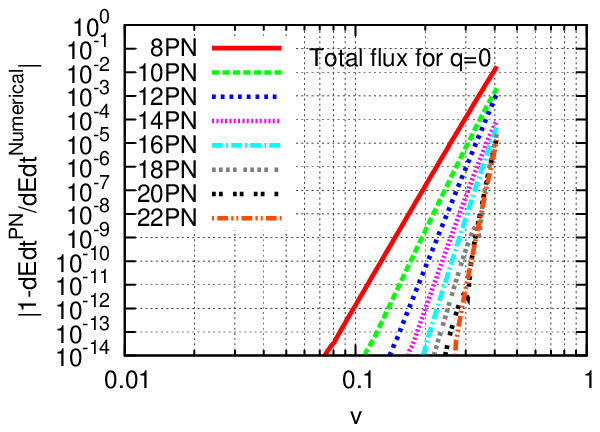}
\end{center}
\caption{(Left) Absolute values of the difference in the energy flux 
down the horizon from numerical results and PN approximation 
as a function of the orbital velocity up to ISCO for $q=0$. 
(Right) Same as the left figure but for the total energy flux, 
which includes fluxes to infinity and the horizon.
}\label{fig:fluxH_sch}
\end{figure}

Figure~\ref{fig:fluxH_sch} shows the relative error in the energy flux 
down the horizon from numerical results and PN approximations 
as a function of the orbital velocity up to ISCO 
in the case of the Schwarzschild black hole. 
The agreement between the numerical energy flux and 
post-Newtonian energy flux becomes better when the PN order is higher 
even around ISCO. The relative error in the 22.5PN energy flux into the horizon 
around ISCO is about $10^{-5}$, which is comparable to the one for 
the 22PN energy flux to infinity in Ref.~\cite{22PN}. 

\subsection{Energy flux: Factorized resummation to PN approximation}
\label{sec:Factorized}
In this section, we compare the total energy flux from 
numerical results with PN results using a factorized resummation. 
The factorized resummation was introduced to improve the convergence in 
the PN energy flux to infinity for a test particle moving 
in Schwarzschild spacetime~\cite{Damour:2007xr,Damour:2007yf,DIN} 
and Kerr spacetime~\cite{PBFRT}. The factorized resummation 
was then extended to the PN energy flux down the horizon of 
the Schwarzschild black hole in Ref.~\cite{NA2012} 
and the Kerr black hole in Ref.~\cite{TBHK2013}. 
\subsubsection{Factorization of the energy flux at infinity}
\label{sec:Factorized8}
In the factorized resummation of the energy flux at infinity, 
we decompose the multipolar gravitational waveforms into five factors as 
\begin{eqnarray}
h_{\ell m}=h_{\ell m}^{({\rm N},\e_p)}\,\hat{S}_{\rm eff}^{(\e_p)}\,T_{\ell m}\,
e^{i\hat \delta_{\ell m}}(\rho_{\ell m})^\ell\,,
\label{eq:rholm8}
\end{eqnarray}
where $\e_p$ denotes the parity of the multipolar waveforms, 
$h_{\ell m}^{({\rm N},\e_p)}$ represents the Newtonian contribution to waveforms, 
$\hat{S}_{\rm eff}^{(\e_p)}$ an effective source term 
for partial waves in the perturbation formalism, 
$T_{\ell m}$ resums the leading logarithms of the tail effects, 
$\hat\delta_{\ell m}$ is the supplemental phase factor and 
$\rho_{\ell m}$ is the $\ell$th root of the amplitude of the waveforms, 
which takes care of a term linear in $\ell$ at 1PN in the waveforms and 
means that a better convergence in the factorized waveforms might be expected 
(for more details see, e.g., Refs.~\cite{Damour:2007xr,Damour:2007yf,DIN}). 

The first factor $h_{\ell m}^{({\rm N},\e_p)}$ takes the form 
\begin{equation}
h_{\ell m}^{({\rm N},\e_p)}=\frac{GM\nu}{c^2\,r}\,n_{\ell m}^{(\e_p)}\,c_{\ell+\e_p}(\nu)\,v^{\ell+\e_p}\,Y^{\ell-\e_p,-m}\,\left(\frac{\pi}{2},\phi\right)\,,
\label{eq:rholmN}
\end{equation}
where $\phi$ is the orbital phase and $n_{\ell m}^{(\epsilon_p)}$ are
\begin{subequations}
\begin{align}
n^{(0)}_{\ell m}=&(im)^\ell\frac{8\pi}{(2\ell+1)!!}\sqrt{\frac{(\ell+1)(\ell+2)}{\ell(\ell-1)}}\,, \\
n^{(1)}_{\ell m}=&-(im)^\ell\frac{16\pi i}{(2\ell+1)!!}\sqrt{\frac{(2\ell+1)(\ell+2)(\ell^2-m^2)}{(2\ell-1)(\ell+1)\ell(\ell-1)}}\,,
\end{align}
\end{subequations}
and $c_{\ell+\e_p}(\nu)$ are functions of the symmetric mass ratio 
$\nu\equiv \mu\,M/(M+\mu)^2$, defined by 
\begin{eqnarray}
c_{\ell+\e_p}(\nu)=\left(\frac{1}{2}-\frac{1}{2}\sqrt{1-4\nu}\right)^{\ell+\e_p-1}+(-)^{\ell+\e_p}\left(\frac{1}{2}+\frac{1}{2}\sqrt{1-4\nu}\right)^{\ell+\e_p-1}.
\end{eqnarray}

The second factor $\hat{S}_{\rm eff}^{(\e_p)}$ in Eq.~(\ref{eq:rholm8}) 
is defined by 
\begin{equation}
\hat{S}_{\rm eff}^{(\e_p)}=\left\{
  \begin{array}{lc}
    \tilde E\, & \,\hbox{for}\,\,\,\e_p=0\,\,(\ell+m={\rm even})\,, \\
    v \tilde L_z/M\, & \hbox{for}\,\,\,\e_p=1\,\,(\ell+m={\rm odd})\,, 
  \end{array}
\right.
\label{eq:rholm8_Slm}
\end{equation}
where $\tilde E$ and $\tilde L_z$ are the specific energy and 
the angular momentum of the particle, given by 
\begin{equation}
\tilde E=\frac{1-2v_r^2+qv_r^3}{\sqrt{1-3v_r^2+2qv_r^3}},\,\,
\tilde L_z=\frac{r_0v_r(1-2qv_r^3+q^2v_r^4)}{\sqrt{1-3v_r^2+2qv_r^3}}.
\label{eq:ELz}
\end{equation}

The third factor $T_{\ell m}$ in Eq.~(\ref{eq:rholm8}) is defined by 
\begin{equation}
T_{\ell m}=\frac{\Gamma(\ell+1-2imM\Omega)}{\Gamma(\ell+1)}\,e^{m\pi M\Omega}\,e^{2i m M\Omega\,\ln(2 m\Omega r_{0s})}\,,
\label{eq:Tlm}
\end{equation}
where $r_{0s}=2M/\sqrt{e}$ is introduced to reproduce the test-particle 
limit waveforms~\cite{PBFRT}. 

The fourth and fifth factors in Eq.~(\ref{eq:rholm8}), 
$\hat\delta_{\ell m}$ and $\rho_{\ell m}$, can be derived by comparing 
the multipolar waveforms Eq.~(\ref{eq:rholm8}) with 
those obtained from Eq.~(\ref{eq:hpm_slm}). 
For the comparison, it is useful to express waveforms Eq.~(\ref{eq:hpm_slm}) 
in terms of the $-2$ spin-weighted spherical harmonics 
$_{-2}Y_{lm}(\theta,\varphi)\equiv _{-2}P_{lm}(\theta)\,e^{im\varphi}/\sqrt{2\pi}$~\cite{PBFRT}
\begin{eqnarray}
h_{+}-i\,h_{\times}&=&-\frac{2}{r}\,\sum _{\ell,m} \frac{\tilde Z^\infty_{\ell m\omega}}{\omega^2}\frac{e^{i m \varphi}}{\sqrt{2\pi}}\,_{-2}S_{\ell m}^{a\omega}(\theta)\,e^{i \omega (r^{*}-t)},\cr
&\equiv&-\frac{2}{r}\,\sum _{l,m} \frac{\tilde C^\infty_{l m\omega}}{\omega^2}\frac{e^{i m \varphi}}{\sqrt{2\pi}}\,_{-2}P_{l m}(\theta)\,e^{i \omega (r^{*}-t)},
\label{eq:hpm_ylm}
\end{eqnarray}
where $_{-2}P_{lm}(\theta)$ is defined as 
\begin{eqnarray}
_{-2}P_{lm}(\theta)&=&(-1)^m\sqrt{\frac{(l+m)!(l-m)!(2l+1)}{2(l+2)!(l-2)!}}\sin^{2l}\left(\frac{\theta}{2}\right)\,\cr
&&\times\sum_{r=0}^{l+2}{{l+2}\choose{r}}{{l-2}\choose{r-2-m}}(-1)^{l-r+2}\cot^{2r-2-m}\left(\frac{\theta}{2}\right)\,.
\end{eqnarray}

From Eq.~\eqref{eq:hpm_ylm}, one can compute $\tilde C^\infty_{l m\omega}$ from 
$\tilde Z^\infty_{\ell m\omega}$
\begin{eqnarray}
\tilde C^\infty_{l m \omega} &=& 
\int_0^{2\pi}\,d\varphi\,\int_0^{\pi}\,\sin\theta\,d\theta\,\sum_{\ell'}\sum_{m'=-\ell'}^{\ell'}\,\tilde Z^\infty_{\ell' m'\omega'}\,\frac{_{-2}S^{a\omega'}_{\ell' m'}(\theta)\,_{-2}P_{l m}(\theta)}{2\pi}\,e^{i\,(m'-m)\,\varphi},\cr 
&=& \int_0^{\pi}\,\sin\theta\,d\theta\,\sum_{\ell'}\,\tilde Z^\infty_{\ell' m\omega}\,{}_{-2}S^{a\omega}_{\ell' m}(\theta)\,_{-2}P_{l m}(\theta),
\label{eq:zlmw2clmw}
\end{eqnarray}
where we used the orthogonality condition of the $-2$ spin-weighted 
spherical harmonics, 
\begin{equation}
\int_0^{2\pi}\,d\varphi\, \int_0^\pi\,\sin\theta\,d\theta\,{}_{-2}Y_{lm}(\theta,\varphi)\,{}_{-2}\bar Y_{l'm'}(\theta,\varphi)=\delta_{ll'}\,\delta_{mm'}\,,
\end{equation}
and $\bar X$ is the complex conjugate of $X$. Note that in 
Eq.~(\ref{eq:zlmw2clmw}) the mixing of $\tilde Z_{\ell m\omega}^{\infty}$ happens 
among the same $m$ and different $\ell$ modes~\cite{PBFRT}. 
Note that the infinite summation over $\ell'$ in Eq.~(\ref{eq:zlmw2clmw})
can be truncated at a certain $\ell'$ for a given post-Newtonian order 
since $\tilde Z_{\ell' m\omega}^{\infty}=O(v^{\ell'+2+\epsilon_p})$ 
(see, e.g., Ref.~\cite{FI2010}). 

Once we obtain $\tilde C^\infty_{\ell m \omega}$ from Eq.~(\ref{eq:zlmw2clmw}), 
it is straightforward to compute $\hat\delta_{\ell m}$ and $\rho_{\ell m}$ 
from the following relation between $h_{\ell m}$ and 
$\tilde C^\infty_{\ell m \omega}$~\cite{BFIS08,FI2010} 
\begin{eqnarray}
h_{\ell m}&=&\int \sin\Theta\,d\Theta\,d\Phi\,(h_{+} - i\,h_{\times})\;_{-2}\bar{Y}_{\ell m}(\Theta, \Phi),\cr
&=&-\frac{2}{r}\,\sum _{\ell',m'}\frac{\tilde C^\infty_{\ell' m'\omega'}\,e^{i m'\Omega r^{*}}}{(m'\Omega)^2}\int \sin\Theta\,d\Theta\,d\Phi\,
e^{-i m'(\Omega\,t-\Phi)}\,_{-2}Y_{\ell' m'}(\Theta,\varphi)\;_{-2}\bar{Y}_{\ell m}(\Theta,\Phi),\cr
&=&-\frac{2}{r}\,\frac{\tilde C^\infty_{\ell m\omega}\,e^{i\,m\Omega (r^{*}-t)}\,e^{i\,m\,\varphi}}{(m\Omega)^2}.
\label{eq:hlm2clmw}
\end{eqnarray}

Using the factorized waveforms $h_{\ell m}$, Eq.~(\ref{eq:rholm8}), 
computed from Eqs.~(\ref{eq:zlmw2clmw}) and (\ref{eq:hlm2clmw}), 
the time-averaged energy flux to infinity is computed as 
\begin{equation}
\left<{dE\over dt}\right>_\infty =
\frac{1}{16\pi}\,\sum_{\ell=2}^{\infty} \sum_{m=-\ell}^{\ell} \,(mM\Omega)^2\,\left\vert {r\over M}\,h_{\ell m}\right\vert^2. 
\label{eq:dEdt8_factorized}
\end{equation}
\subsubsection{Factorization of the energy flux down the horizon}
\label{sec:FactorizedH}
For the factorized resummation of the energy flux down the horizon, 
the modal energy flux is decomposed as~\cite{NA2012,TBHK2013}
\begin{equation}
\eta_{\ell m}^{\rm H} = \left(1-\frac{2 v^3 r_{+}}{a}\right)\,\eta_{\ell m}^{\rm N,H}
\,(\hat{S}_{\rm eff}^{(\e_p)})^2\,(\rho_{\ell m}^{\rm H})^{2 \ell},
\label{eq:etaH_factorized}
\end{equation}
where the factor $(1-2 v^3 r_{+}/a)$ is motivated by the factor 
$k=\omega - ma/(2Mr_+)=m\,(\Omega-a/(2Mr_+))$ in Eq.~(\ref{eq:alphaH}), 
which is responsible for the sign of the modal energy flux to the horizon. 

The second factor $\eta_{\ell m}^{\rm N,H}$ represents the leading term 
in the modal energy flux into the horizon and takes the form 
\begin{equation}
\eta_{\ell m}^{\rm N,H} = v^{4(\ell-2) +2\e_p}\,n^{({\rm H},\e_p)}_{\ell m}\,c^{\rm H}_{\ell m}(q)\,,
\label{eq:etaH_leading}
\end{equation}
where
\begin{subequations}
\begin{align}
n^{({\rm H},0)}_{\ell m}=&-\frac{5}{32}\frac{(\ell+1)(\ell+2)}{\ell(\ell-1)}\frac{2\ell+1}{[(2\ell+1)!!]^2}\,\frac{(\ell-m)!}{[(\ell-m)!!]^2} \frac{(\ell+m)!}{[(\ell+m)!!]^2} \,,\cr
n^{({\rm H},1)}_{\ell m}=&-\frac{5}{8\ell^2}\frac{(\ell+1)(\ell+2)}{\ell(\ell-1)}\frac{2\ell+1}{[(2\ell+1)!!]^2}\,\frac{[(\ell-m)!!]^2}{(\ell-m)!}\frac{[(\ell+m)!!]^2}{(\ell+m)!} \,,
\end{align}
\end{subequations}
and $c_{\ell+\e_p}(\nu)$ 
\begin{eqnarray}
c^{\rm H}_{\ell m}(q) &=& \frac{1}{q}\prod_{k=0}^{\ell}{\left[k^2 + \left(m^2 - k^2\right) q^2\right]},\cr
&=&q\,m^2\,\left(1 - q^2\right)^{\ell}\,\left(1 - \frac{i m q}{\sqrt{1 - q^2}}\right)_{\ell}\,\left(1 + \frac{i m q}{\sqrt{1 - q^2}}\right)_{\ell} \,,
\end{eqnarray}
where $(z)_n=\Gamma(z+n)/\Gamma(z)$. 

The definition for the third factor $\hat{S}_{\rm eff}^{(\e_p)}$ is the same as 
in Eq.~(\ref{eq:rholm8_Slm}), which is used for the resummed multipolar 
waveforms Eq.~(\ref{eq:rholm8}). The fourth factor $\rho_{\ell m}^{\rm H}$ is 
the $2\ell$th root of the residual amplitude of the modal energy flux 
and can be derived by comparing the Taylor expanded modal energy flux 
$\eta_{\ell m}^{\rm H}$ with the factorized modal energy flux 
Eq.~(\ref{eq:etaH_factorized}). 
\subsubsection{Comparisons with numerical results}
\label{sec:Factorized_comparison}
\begin{figure}[t]
\begin{center}
\includegraphics[width=69mm]{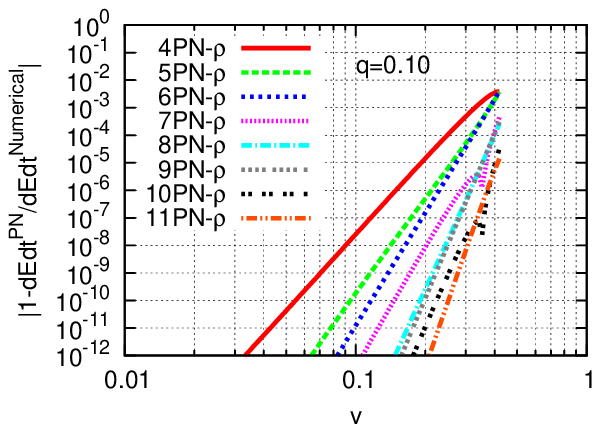}%
\includegraphics[width=69mm]{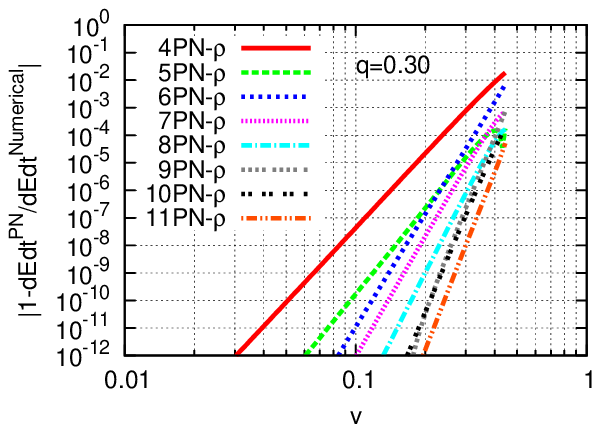}\\
\includegraphics[width=69mm]{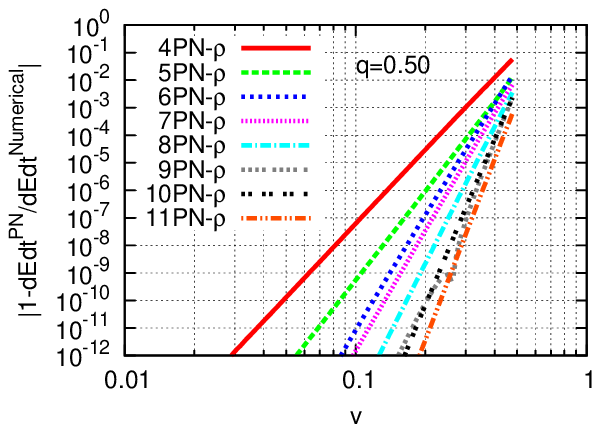}%
\includegraphics[width=69mm]{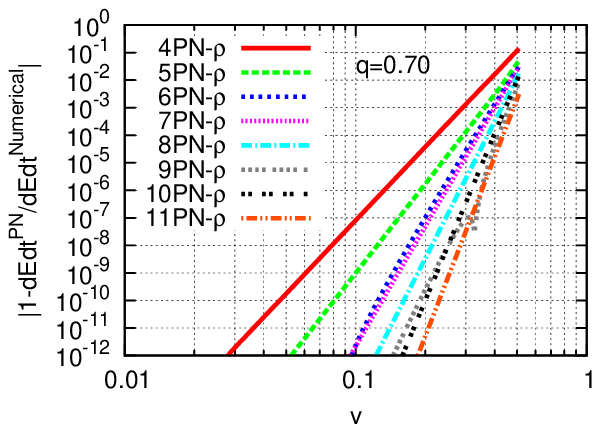}\\
\includegraphics[width=69mm]{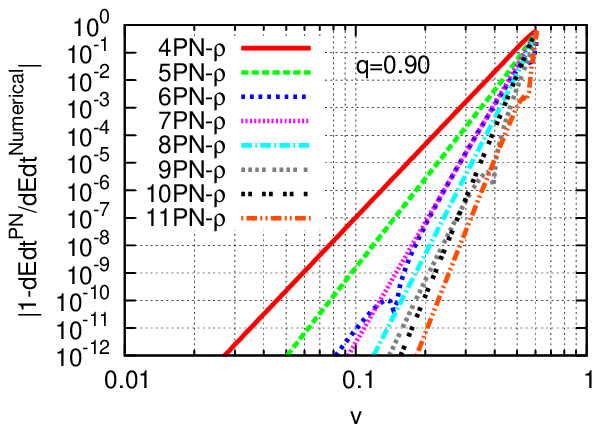}
\includegraphics[width=69mm]{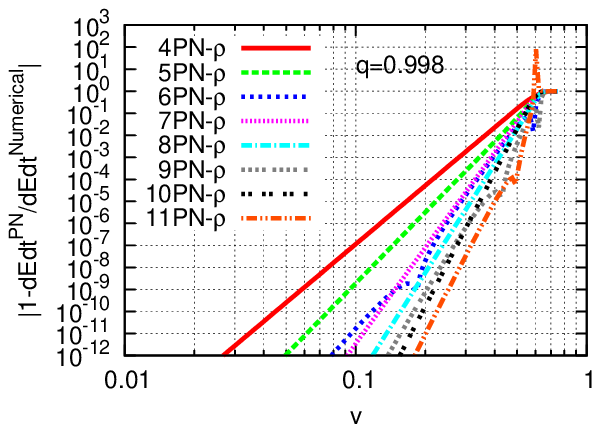}%
\end{center}
\caption{Same as Fig.~\ref{fig:flux_taylor_1} but using 
factorized resummation to the energy flux in the post-Newtonian approximation. 
The relative error for 11PN is less than $10^{-5}$ when $v\lessapprox 0.4$, 
whose region is larger than $v\lessapprox 0.33$ 
for the Taylor expanded energy flux  
in Fig.~\ref{fig:flux_taylor_1}. 
}\label{fig:flux_factorized_1}
\end{figure}

\begin{figure}[t]
\begin{center}
\includegraphics[width=69mm]{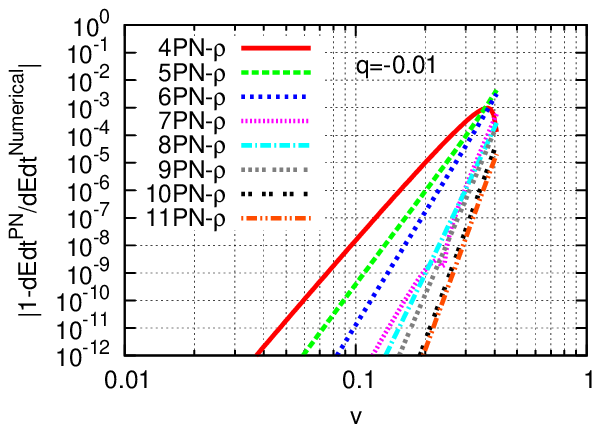}%
\includegraphics[width=69mm]{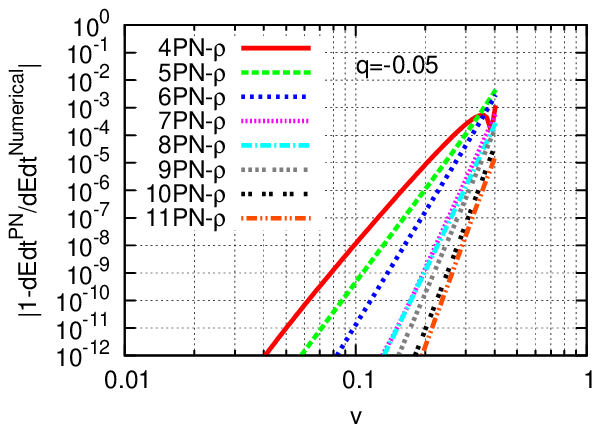}\\
\includegraphics[width=69mm]{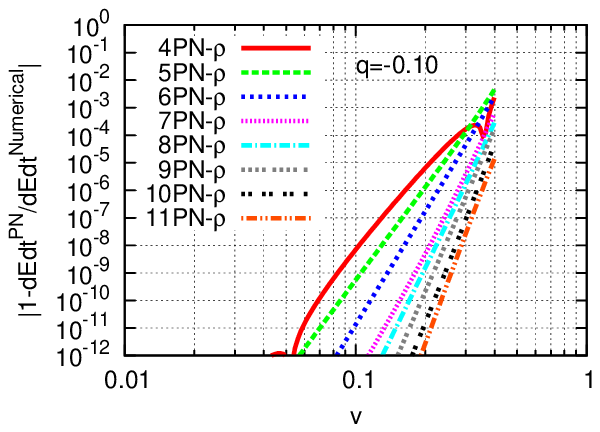}%
\includegraphics[width=69mm]{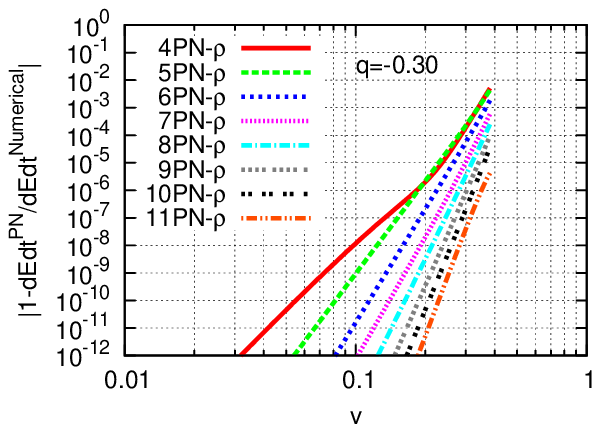}\\
\includegraphics[width=69mm]{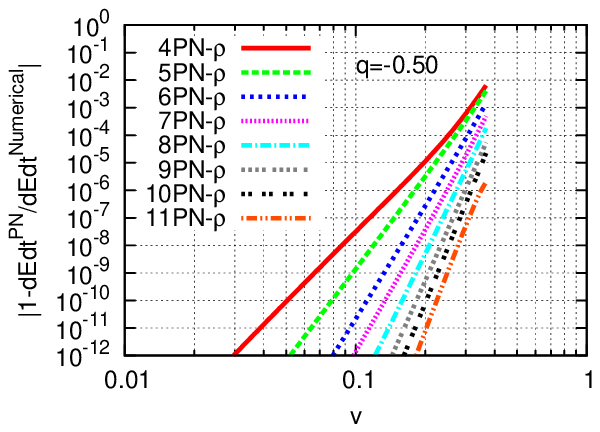}%
\includegraphics[width=69mm]{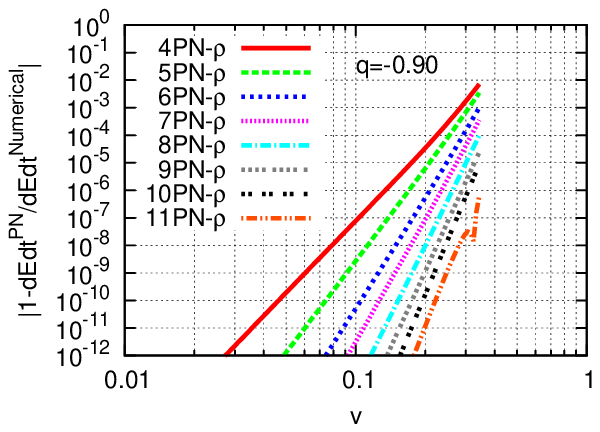}
\end{center}
\caption{Same as Fig.~\ref{fig:flux_taylor_2} but using 
factorized resummation to the energy flux in the post-Newtonian approximation. 
The relative error for 11PN is less than $10^{-5}$ when $v\lessapprox 0.4$, 
whose region is larger than $v\lessapprox 0.33$ 
for the Taylor expanded energy flux  
in Fig.~\ref{fig:flux_taylor_2}. 
}\label{fig:flux_factorized_2}
\end{figure}

Figures~\ref{fig:flux_factorized_1} and \ref{fig:flux_factorized_2} show 
the relative error in the total energy flux from numerical results and 
PN approximations as a function of the orbital velocity up to ISCO 
using the factorized resummation to PN approximations~\cite{DIN}. 
From these figures, one will find that 
the relative error becomes smaller as PN order becomes higher
for $v\le 0.3$, 
except for accidental agreements for certain values of the velocity. 
However, the relative error around ISCO does not necessarily become smaller 
for higher PN orders when $q > 0.3$. 
The relative error for 11PN is smaller than $10^{-5}$ when $v\lessapprox 0.4$, 
irrespective of the values of $q$ investigated in the paper. 
We note that the region of the velocity, $v\lessapprox 0.4$, 
is larger than the one using the Taylor expanded PN energy flux, 
$v\lessapprox 0.33$. 

\subsection{Energy flux: Exponential resummation to PN approximation}
\label{sec:Exponential}
In this section, we compare the total energy flux from numerical results with 
PN results using the exponential resummation~\cite{Isoyama:2012bx,Johnson2014}.

In the exponential resummation, the modal energy fluxes to infinity, 
$\eta_{\ell m}^\infty$, and the horizon, $\eta_{\ell m}^{\rm H}$, are decomposed as 
\begin{eqnarray}
\eta_{\ell m}^{\infty} = {1\over {1-3v_r^2+2qv_r^3}}\,\eta_{\ell m}^{{\rm N},\infty}\,\exp\left[ \ln\left(\hat\eta_{\ell m}^{\infty}\right)\right],\,\,\,
\eta_{\ell m}^{\rm H} = \frac{1-\frac{2 v^3 r_{+}}{a}}{1-3v_r^2+2qv_r^3}\,\eta_{\ell m}^{{\rm N},{\rm H}}\,\exp\left[ \ln\left(\hat\eta_{\ell m}^{\rm H}\right)\right],
\label{eq:eta8H_exp}
\end{eqnarray}
where $\eta_{\ell m}^{{\rm N},\infty}$ and $\eta_{\ell m}^{{\rm N},{\rm H}}$ are 
the leading terms for $\eta_{\ell m}^{\infty}$ and $\eta_{\ell m}^{{\rm H}}$ 
respectively, the denominator $(1-3v_r^2+2qv_r^3)$ is 
the square of the denominator 
of $\hat{S}_{\rm eff}^{(\e_p)}$ in Eq.~(\ref{eq:rholm8}), and 
the factor $(1-2 v^3 r_{+}/a)$ is motivated by the factor 
$k=\omega - ma/(2Mr_+)=m/M\,(v^3-a/(2r_+))$ in Eq.~(\ref{eq:alphaH}), 
which is again responsible for the sign of the modal energy flux 
to the horizon. 

Similarly to $\eta_{\ell m}^{{\rm N},{\rm H}}$ defined in 
Eq.~(\ref{eq:etaH_leading}), the explicit expression for 
$\eta_{\ell m}^{{\rm N},\infty}$ can be derived from factors 
$n_{\ell m}^{(\epsilon_p)}$ and $c_{\ell+\e_p}(\nu)$ 
in the Newtonian contribution to waveforms $h_{\ell m}^{({\rm N},\e_p)}$, 
Eq.~(\ref{eq:rholmN}), as 
\begin{equation}
\eta_{\ell m}^{{\rm N},\infty}={5\over 256\pi^2}\,m^2\,
\left|n_{\ell m}^{(\epsilon_p)}\right|^2\,\left(c_{\ell+\e_p}(\nu)\right)^2\,
v^{2(\ell-2)+2\e_p}\,\left(P_{\ell-\e_p,-m}\left({\pi\over 2}\right)\right)^2. 
\end{equation}

The factors $\hat\eta_{\ell m}^{\infty}$ and $\hat\eta_{\ell m}^{{\rm H}}$ 
in the exponential resummation, Eq.~(\ref{eq:eta8H_exp}), 
can be derived by comparing with the Taylor expanded modal energy fluxes 
$\eta_{\ell m}^{\infty}$ and $\eta_{\ell m}^{{\rm H}}$. 

\begin{figure}[t]
\begin{center}
\includegraphics[width=69mm]{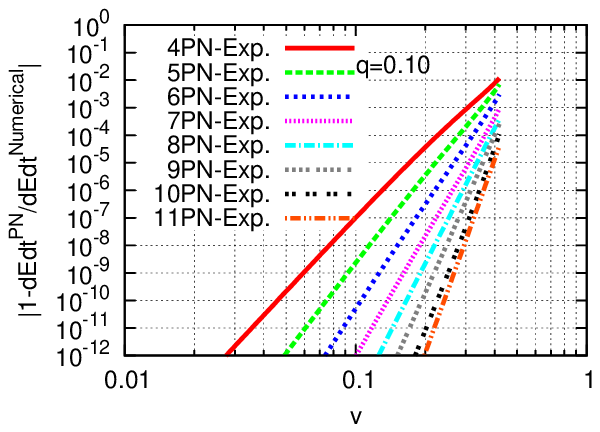}%
\includegraphics[width=69mm]{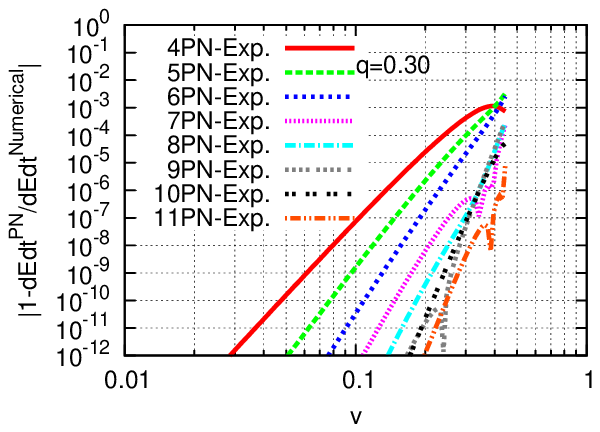}\\
\includegraphics[width=69mm]{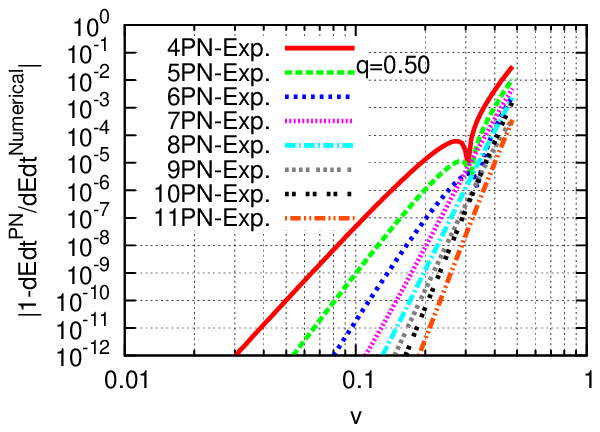}%
\includegraphics[width=69mm]{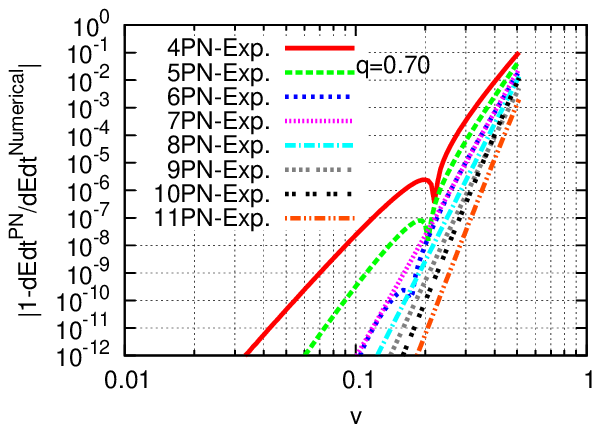}\\
\includegraphics[width=69mm]{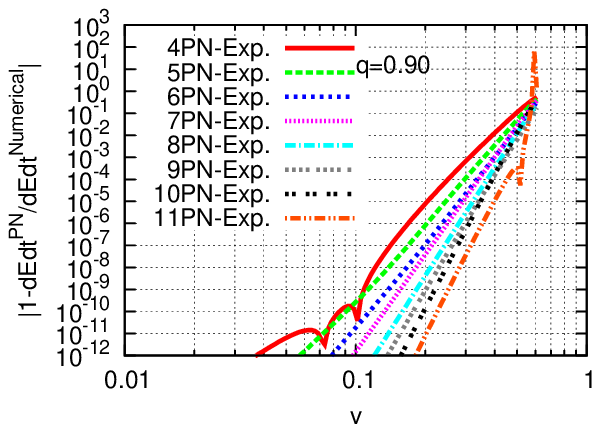}
\includegraphics[width=69mm]{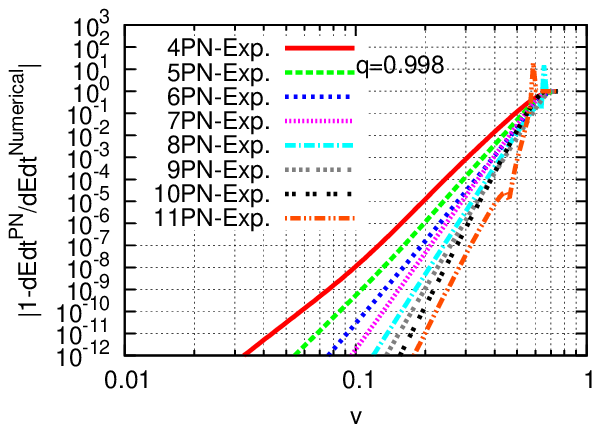}%
\end{center}
\caption{Same as Fig.~\ref{fig:flux_taylor_1} but using 
exponential resummation to the energy flux in the post-Newtonian approximation. 
The relative error for 11PN is less than $10^{-5}$ when $v\lessapprox 0.4$, 
whose region is larger than $v\lessapprox 0.33$ 
for the Taylor expanded energy flux  
in Fig.~\ref{fig:flux_taylor_1}. 
}\label{fig:flux_exponentioal_1}
\end{figure}

\begin{figure}[t]
\begin{center}
\includegraphics[width=69mm]{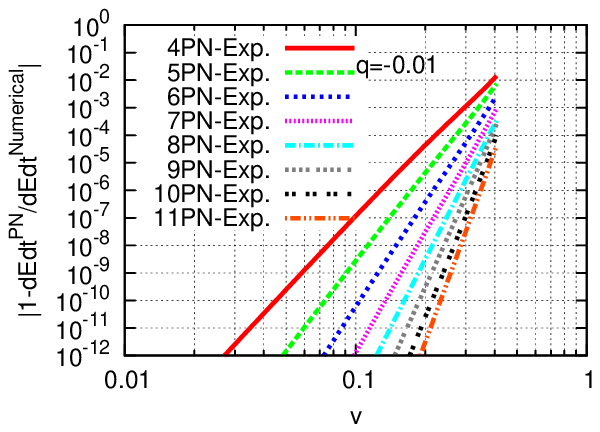}%
\includegraphics[width=69mm]{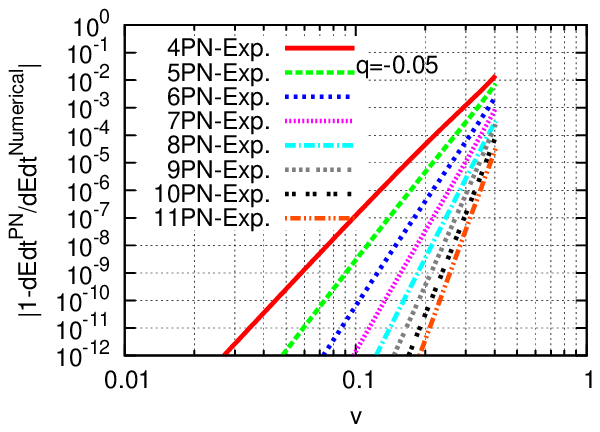}\\
\includegraphics[width=69mm]{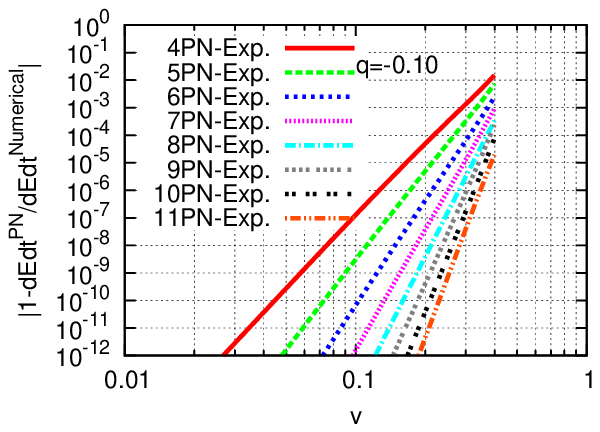}%
\includegraphics[width=69mm]{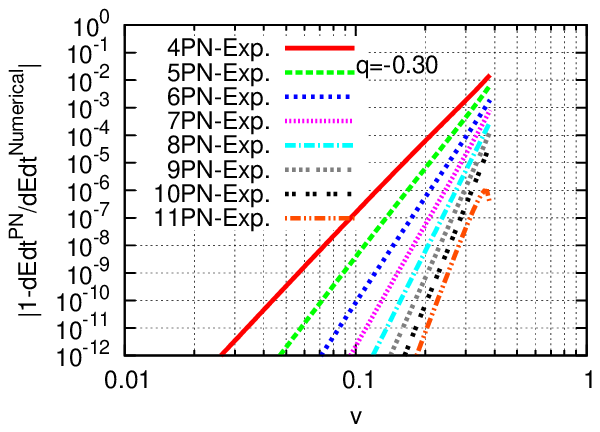}\\
\includegraphics[width=69mm]{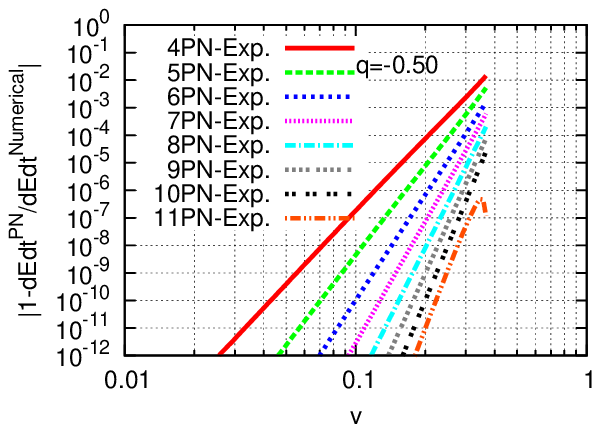}%
\includegraphics[width=69mm]{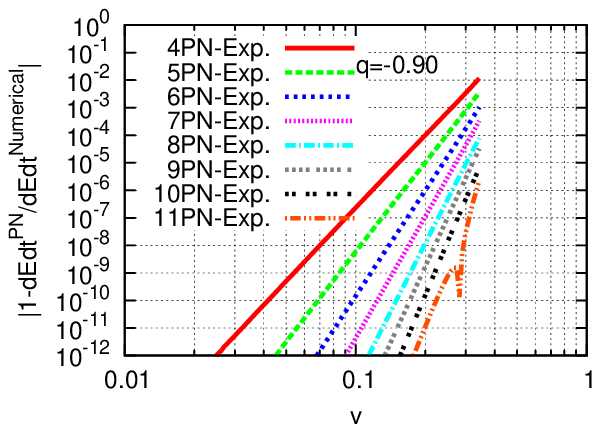}
\end{center}
\caption{Same as Fig.~\ref{fig:flux_taylor_2} but using 
exponential resummation to the energy flux in the post-Newtonian approximation. 
The relative error for 11PN is less than $10^{-5}$ when $v\lessapprox 0.4$, 
whose region is larger than $v\lessapprox 0.33$ 
for the Taylor expanded energy flux  
in Fig.~\ref{fig:flux_taylor_2}. 
}\label{fig:flux_exponentioal_2}
\end{figure}

Figures~\ref{fig:flux_exponentioal_1} and \ref{fig:flux_exponentioal_2} show 
the relative error in the total energy flux from numerical results and 
PN approximations as a function of the orbital velocity up to ISCO 
using exponential resummation to PN approximations~\cite{Isoyama:2012bx}. 
From these figures, one will find that 
the relative error becomes smaller as the PN order becomes higher
for $v\le 0.3$, 
except for accidental agreements for certain values of the velocity. 
However, the relative error around ISCO does not necessarily become smaller 
at higher PN orders when $q > 0.3$. 
The relative error for 11PN is smaller than $10^{-5}$ when $v\lessapprox 0.4$, 
irrespective of values of $q$ investigated in the paper. 
Again, we note that the region of the velocity, $v\lessapprox 0.4$, 
is larger than the one using the Taylor expanded PN energy flux, 
$v\lessapprox 0.33$. 

\subsection{Phase difference during the two-year inspiral}
\label{sec:dephase}
We compare the orbital phase from PN results with numerical results 
during two-year inspirals to estimate the applicability of the PN results 
in the data analysis. 
For the comparison, 
we choose two representative systems of EMRIs in the eLISA frequency band, 
System-I and System-II, following Refs.~\cite{EOB_EMRI,YBHPBMT,14PN,22PN}. 
System-I is an early inspiral of an EMRI with masses 
$(M,\m)=(10^5,10)M_{\odot}$, i.e. $\m/M=10^{-4}$, which reaches $r_0\simeq 16M$ 
after the two-year inspiral. 
System-II is a late inspiral of an EMRI with masses 
$(M,\m)=(10^6,10)M_{\odot}$, i.e. $\m/M=10^{-5}$, which reaches ISCO 
after the two-year inspiral. 
Although the initial values for orbital radius, velocity, and 
GW frequency depend on the spin of the Kerr black hole, 
System-I inspirals from $r_0\simeq 29M$ to $r_0\simeq 16M$ 
with associated velocities $v\in [0.2,0.25]$ and frequencies 
$f_{\rm GW}\in [4\times 10^{-3},10^{-2}]$Hz, 
while System-II explores orbital separation 
in the range $r_0/M\in [r_{\rm ISCO},11M]$, 
velocities $v\in [0.3,v_{\rm ISCO}]$ and frequencies 
$f_{\rm GW}\in [10^{-3},f_{\rm GW}^{\rm ISCO}]$Hz. 
The orbital phase for the System-I (System-II) after the two-year inspiral 
is about $10^6$ ($5\times 10^5$) rad. 
Moreover, System-I (System-II) sweeps 
the high- (low-) frequency region of the eLISA frequency band. 

For the calculation of the orbital phase, we define the phase as 
$\Psi_{\ell m}(t)=m\,\int_0^t\,\Omega(t')dt'$, where 
$\Omega(t)=M^{1/2}/r(t)^{3/2}/(1+qM^{3/2}/r(t)^{3/2})$ 
is the angular frequency of the particle and $r(t)$ is the orbital radius 
as a function of time. The orbital radius $r(t)$ is derived as 
$r(t)=\int^t (dr/dt')\,dt'=\int^t (\partial r/\partial\tilde E)\,(d\tilde E/dt')\,dt'$, where $\langle d\tilde E/dt\rangle$ is computed by 
the energy balance equation, $\langle d\tilde E/dt\rangle=-\langle dE/dt\rangle_\infty-\langle dE/dt\rangle_{\rm H}$. To save computation time, 
we apply cubic spline interpolation~\cite{Recipes} to perform the integration 
using $10^3$ data points for $(v,\langle d\tilde E/dt\rangle)$, 
i.e. $(v,dr/dt)$, in the range from $v=0.01$ to 
$v=v_{\rm ISCO}$~\cite{Hughes2001,EOB_EMRI,14PN,22PN}. 
The computation time to perform the numerical integration is less than 
a second if we use the cubic spline interpolation. 

\begin{figure}[htbp]
\begin{center}
\includegraphics[width=69mm]{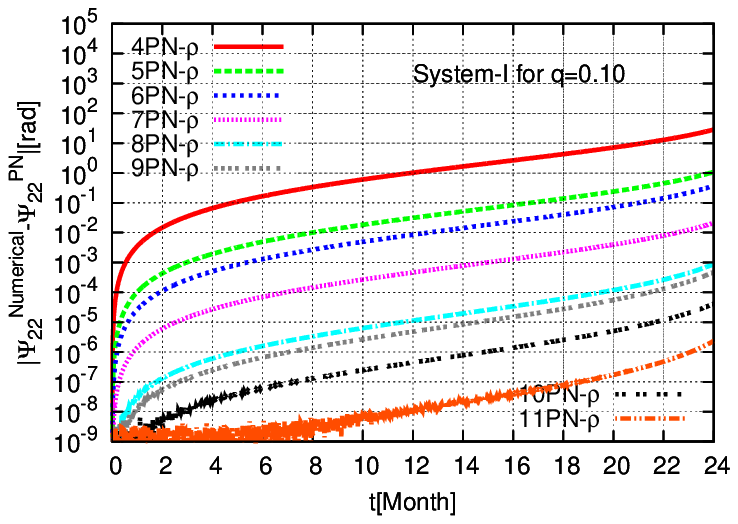}%
\includegraphics[width=69mm]{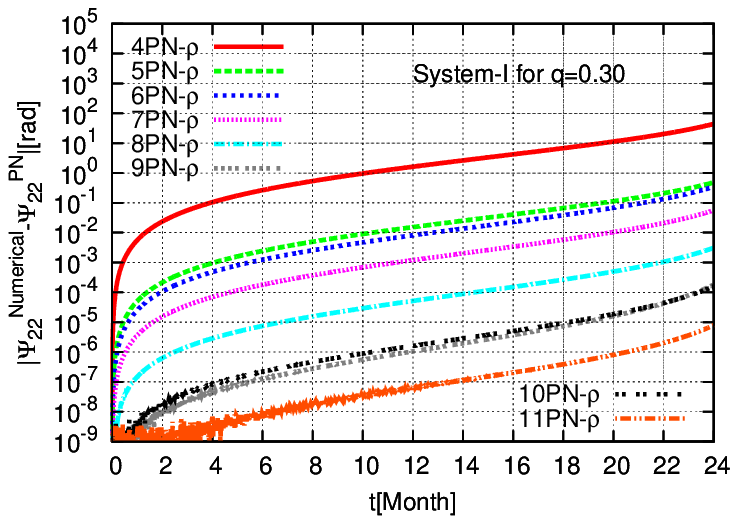}\\
\includegraphics[width=69mm]{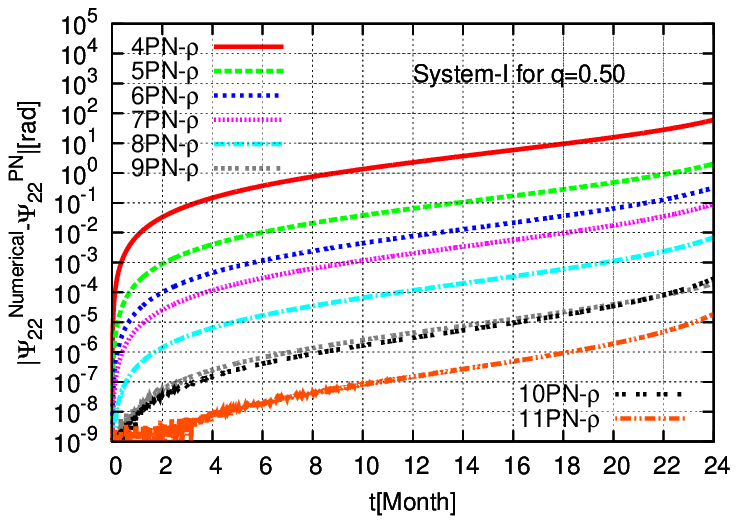}%
\includegraphics[width=69mm]{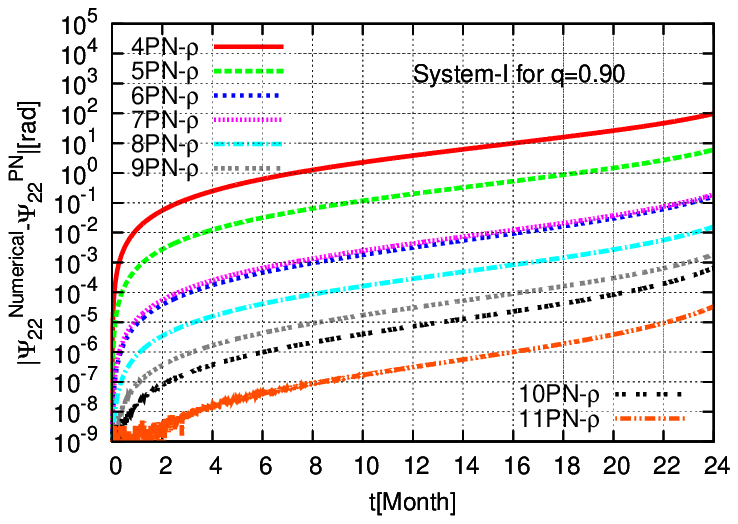}
\end{center}
\caption{Absolute values of the dephasing during the two-year inspiral 
between the factorized PN and the numerical results 
for the dominant $\ell=m=2$ mode 
as a function of time in months when $q=0.1,\,0.3,\,0.5$ and $0.9$. 
These panels show the dephases for System-I with masses 
$(M,\m)=(10^5,10)M_{\odot}$, which inspirals from 
$r_0\simeq 29M$ to $r_0\simeq 16M$ with associated frequencies 
$f_{\rm GW}\in [4\times 10^{-3},10^{-2}]$Hz. 
These inspirals represent the early inspiral phase in the eLISA band. 
The dephases between the 11PN results and numerical results 
after the two-year inspiral are less than $10^{-4}$ rad. 
}\label{fig:dephase_M5}
\end{figure}

\begin{figure}[htbp]
\begin{center}
\includegraphics[width=69mm]{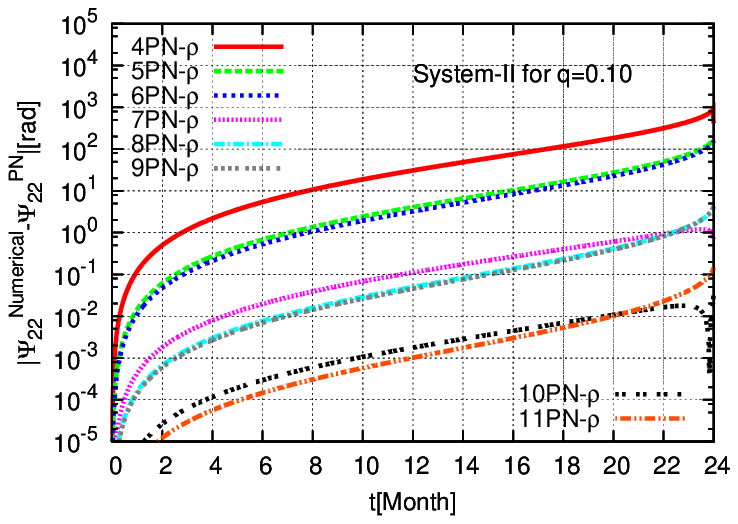}%
\includegraphics[width=69mm]{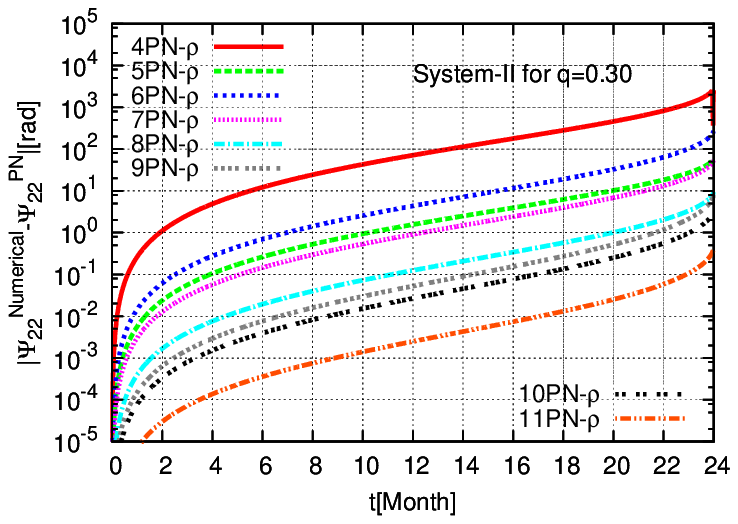}\\
\includegraphics[width=69mm]{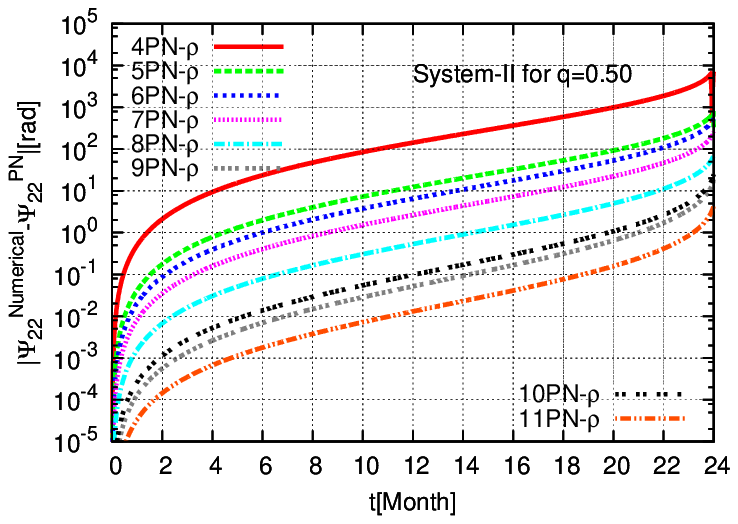}
\includegraphics[width=69mm]{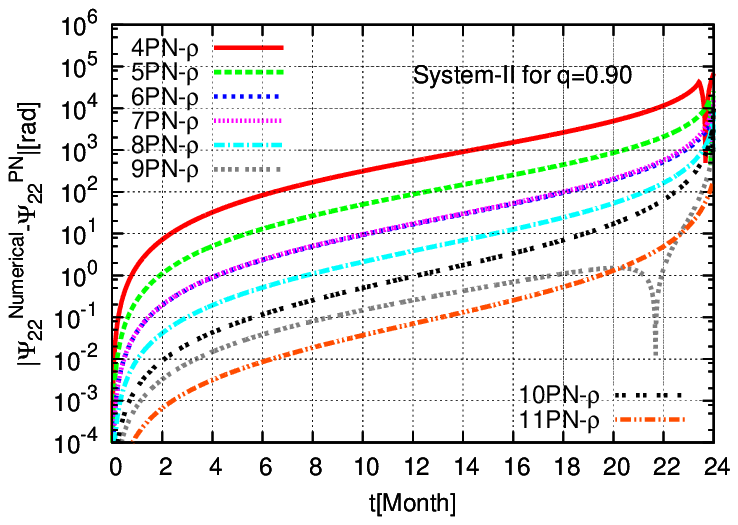}
\end{center}
\caption{Same as Fig.~\ref{fig:dephase_M5} but for System-II 
with masses $(M,\m)=(10^6,10)M_{\odot}$, 
orbital radius in the range $r_0/M\in [r_{\rm ISCO},11]$ and 
frequencies in the range $f_{\rm GW}\in [10^{-3},f_{\rm GW}^{\rm ISCO}]$Hz. 
These inspirals represent the late inspiral phase in the eLISA band. 
The dephases between the 11PN results and numerical results 
for $q\le 0.3$ ($q>0.3$) after the two-year inspiral are less (larger) than 
a radian. 
}\label{fig:dephase_M6}
\end{figure}

Figures~\ref{fig:dephase_M5} and \ref{fig:dephase_M6} show 
absolute values of the difference in the orbital phase for 
the dominant $\ell=m=2$ mode between the PN and the numerical results 
during two-year inspirals for several values of the spin of the black hole. 
As for the PN approximations, we show results 
using the factorized resummation in Sec.~\ref{sec:Factorized}, 
which are better than those using the Taylor expanded PN energy flux 
and comparable to those using the exponential resummation. 
The dephases between the 11PN results and numerical results 
after the two-year inspiral are less than $10^{-4}$ rad for System-I. 
However, the dephases after the two-year inspiral become larger than 
a radian for System-II when $q>0.3$. Thus, one has to derive higher PN order 
results for the energy flux to achieve a dephase of less than a radian 
for System-II when $q>0.3$, 
which represents a stronger-field situation than the one for System-I. 

\section{Summary}
\label{sec:summary}
We have investigated gravitational waves from a particle moving in 
circular orbits in Kerr spacetime using the post-Newtonian approximation 
and computed the energy flux up to 11PN. 
We have also computed the energy flux down the event horizon for a particle 
in circular orbits around a Schwarzschild black hole at 22.5PN 
beyond the Newtonian approximation to fill the gap in 
the PN order between the energy flux at infinity, currently known at 22PN, 
and the event horizon, previously known at 6.5PN 
beyond Newtonian approximation. 

To investigate how higher PN order expressions improve the applicability to 
data analysis of eLISA/NGO, comparisons between PN results and high-precision 
numerical results in black hole perturbation theory have been done. 
We first compared PN energy flux to numerical energy flux 
and found that the region of validity in the PN energy flux becomes larger 
as the PN order becomes higher. If the relative error of the energy flux 
in the PN approximation should be less than $10^{-5}$, the energy flux at 11PN 
satisfies this requirement for $v\lessapprox 0.33$, 
which clearly shows an improvement 
from $v\lessapprox 0.13$ in an earlier work at 4PN~\cite{TSTS}. 
The region of validity in the 11PN energy flux can become larger, 
$v\lessapprox 0.4$, if one uses resummation techniques such as 
factorized resummation~\cite{DIN} and 
exponential resummation~\cite{Isoyama:2012bx}. 
\footnote{In Ref.~\cite{Shah_20PN}, the relative error in the fitting formula of the 20PN energy flux using a series expansion in $q$ is less than $10^{-4}$ for $\Omega\lessapprox 0.8\Omega_{\rm ISCO}$ when $q=0.5$ and $\Omega\lessapprox 0.35\Omega_{\rm ISCO}$ when $q=0.9$, where $\Omega_{\rm ISCO}$ is the angular frequency of the particle at ISCO. The region of $\Omega$ in Ref.~\cite{Shah_20PN} is comparable to the one for our 11PN results using resummation techniques, which is estimated as $\Omega\lessapprox 0.78\Omega_{\rm ISCO}$ when $q=0.5$ and $\Omega\lessapprox 0.37\Omega_{\rm ISCO}$ when $q=0.9$.} 
The region of validity might become further larger if one takes account of 
the structure of homogeneous solutions of the Teukolsky equation 
more carefully~\cite{Johnson2014}.

Finally, we compared the orbital phase during the two-year inspiral 
using the factorized resummed PN flux and the high-precision numerical flux. 
We found that the dephase is less than 1 ($10^{-4}$) rad 
for late (early) inspirals when $q\le 0.3$ ($q\le 0.9$). 
This implies that the 11PN factorized  resummed flux may be used to detect 
early inspirals in the data analysis of eLISA/NGO. 
To detect gravitational waves from late inspirals when $q> 0.3$, however, 
it is necessary to obtain higher PN order expressions than 11PN. 
From numerical calculations in black hole perturbation theory, 
it is estimated that we may need to compute at least up to $\ell=30$, 
i.e. 28PN, to obtain the relative error of $10^{-5}$ in the energy flux 
at ISCO for $q=0.9$~\cite{22PN}. 
If it is not possible to perform such a high PN order calculation, 
it may be necessary to use other approaches that compute 
unknown PN coefficients by numerical fitting
~\cite{EOB_EMRI,YBHPBMT,Shah_20PN}. 
\section*{Acknowledgements}
It is our pleasure to thank Bala Iyer for his continuous encouragement 
and useful comments on the manuscript. We also thank Abhay Shah for 
sharing his results before submitting Ref.~\cite{Shah_20PN}, 
which was very helpful for correcting errors in our results. 
This work was supported by 
the European Union FEDER funds, 
the Spanish Ministry of Economy and Competitiveness (Projects No. 
FPA2010-16495 and No. CSD2007-00042), the Conselleria 
d'Economia Hisenda i Innovacio of the Govern de les Illes Balears
and 
the European Union's FP7 ERC Starting Grant ``The dynamics of black holes:
testing the limits of Einstein's theory'' grant agreement no. DyBHo--256667. 
Some analytic calculations were carried out on HA8000/RS440 at 
Yukawa Institute for Theoretical Physics in Kyoto University. 
\appendix
\section{Source term of the Teukolsky equation}
\label{sec:source}.
$A_{nn0}$ and etc. in Eq.~(\ref{eq:source}) are defined as 
\begin{align}
A_{n\,n\,0}&={-2 \over \sqrt{2\pi}\Delta^2}
C_{n\,n}\,\rho^{-2}\,{\overline \rho}^{-1}
\mathcal{L}^{\dagger}_1\,\{\rho^{-4}\mathcal{L}^{\dagger}_2\,(\rho^3 {}_{-2}S_{\ell m}^{a\,\omega}(\theta))\},\cr
A_{{\overline m}\,n\,0}&={2 \over \sqrt{\pi}\Delta} C_{{\overline m}\,n}\,\rho^{-3} \Bigl[\left(\mathcal{L}^{\dagger}_2\,{}_{-2}S_{\ell m}^{a\,\omega}(\theta)\right) \Bigl({iK \over \Delta}+\rho+{\overline \rho}\Bigr) -a\sin\theta \,{}_{-2}S_{\ell m}^{a\,\omega}(\theta)\,{K \over \Delta} ({\overline \rho}-\rho)\Bigr],\cr
A_{{\overline m}\,{\overline m}\,0}&=-{1 \over \sqrt{2\pi}}\rho^{-3}\,{\overline \rho}\,C_{{\overline m}\,{\overline m}}\,{}_{-2}S_{\ell m}^{a\,\omega}(\theta)\Bigl[-i\Bigl({K \over \Delta}\Bigr)_{,r}-{K^2 \over \Delta^2}+2i\rho {K \over \Delta}\Bigr],\cr
A_{{\overline m}\,n\,1}&={2\over \sqrt{\pi}\Delta }\rho^{-3}\,C_{{\overline m}\,n}[\mathcal{L}^{\dagger}_2\,{}_{-2}S_{\ell m}^{a\,\omega}(\theta)+ia\sin\theta({\overline \rho}-\rho){}_{-2}S_{\ell m}^{a\,\omega}(\theta)],\cr
A_{{\overline m}\,{\overline m}\,1}&=-{2 \over \sqrt{2\pi}}\rho^{-3}\,{\overline \rho}\,C_{{\overline m}\,{\overline m}}\,{}_{-2}S_{\ell m}^{a\,\omega}(\theta)\Bigl(i{K \over \Delta}+\rho\Bigr),\cr
A_{{\overline m}\,{\overline m}\,2}&=-{1\over \sqrt{2\pi}}\rho^{-3}\,{\overline \rho}\,C_{{\overline m}\,{\overline m}}\,{}_{-2}S_{\ell m}^{a\,\omega}(\theta), 
\label{eq:ann} 
\end{align}
where $\mathcal{L}^{\dagger}_\sigma=\partial_\theta-{m}/{\sin\theta}+a\omega \sin\theta +\sigma\cot\theta$, $\rho=1/(r-i a \cos\theta)$ and 
\begin{align}
C_{n\,n}&={1\over 4\Sigma^3 \dot t}\left[\tilde E(r^2+a^2)-a\tilde L_z+\Sigma{dr\over d\tau} \right]^2,\cr
C_{{\overline m}\,n}&=-{\rho \over 2\sqrt{2}\Sigma^2 \dot t}\left[\tilde E(r^2+a^2)-a\tilde L_z
+\Sigma{dr\over d\tau} \right]\left[i\sin\theta\Bigl(a\tilde E-{\tilde L_z \over \sin^2\theta}\Bigr)\right], \cr
C_{{\overline m}\,{\overline m}}&={\rho^2 \over 2\Sigma \dot t }\left[i\sin\theta \Bigl(a\tilde E-{\tilde L_z \over \sin^2\theta}\Bigr)\right]^2,
\label{eq:cnn}
\end{align}
with $\dot t=dt/d\tau$ and $\Sigma=r^2+a^2\cos^2\theta$.
\section{Spin-weighted spheroidal harmonics}
\label{sec:PN_Sph}
Using $x = \cos\theta$, the angular Teukolsky equation (\ref{eq:Sph_costh}) 
takes the form 
\begin{equation}
\left[(1-x^{2})\frac{d^2}{dx^{2}}-2x\frac{d}{dx}+{\xi}^{2} x^{2}-\frac{m^{2}+s^{2}+2msx}{1-x^{2}}-2s\xi x+ {}_{s}E_{\ell m}(\xi) \right]\ {}_{s}S_{\ell m}^{a\omega}(x)=0,
\label{eq:Sph_x}
\end{equation}
where $\xi = a\,\omega$ and ${}_{s}E_{\ell m}(\xi)=\lambda+s(s+1)-a^{2} \omega^{2}+2\,a\,m\,\omega$.

When $\xi=0$, the solutions ${}_{s}S_{\ell m}^{a\omega}(x)$ in Eq.~(\ref{eq:Sph_x}) 
reduce to the spin-weighted spherical harmonics and 
the eigenvalue $_{s}E_{\ell m}(\xi)$ becomes $\ell(\ell+1)$ \cite{TP1974}. 
Thus, it might be useful to express the spin-weighted spheroidal 
harmonics in a series of the spin-weighted spherical 
harmonics~\cite{PT1973,Hughes2000,TSTS,chapter}. 

Taking account of singularities at $x=\pm 1$ and $\infty$
in the differential equation (\ref{eq:Sph_x}), it is also possible 
to expand the spin-weighted spheroidal harmonics in a series of 
Jacobi polynomials~\cite{Fackerell,FT1}. 
For this purpose, we introduce new functions ${}_{s}U_{\ell m}(x)$ 
and ${}_{s}V_{\ell m}(x)$ through 
\begin{eqnarray}
{}_{s}S_{\ell m}^{a\omega}(x)=e^{\xi x}\left(\frac{1-x}{2}\right)^{\frac{\alpha}{2}}
\left(\frac{1+x}{2}\right)^{\frac{\beta}{2}}\, {}_{s}U_{\ell m}(x),
\label{eq:SpheU}
\end{eqnarray}
and
\begin{eqnarray}
{}_{s}S_{\ell m}^{a\omega}(x)=e^{-\xi x}\left(\frac{1-x}{2}\right)^{\frac{\alpha}{2}}
\left(\frac{1+x}{2}\right)^{\frac{\beta}{2}}\, {}_{s}V_{\ell m}(x),
\label{eq:SpheV}
\end{eqnarray}
where $\alpha = |m+s|$ and $\beta = |m-s|$. 
Note that Eqs. (\ref{eq:SpheU}) and (\ref{eq:SpheV}) imply 
\begin{eqnarray}
\,{}_{s}V_{\ell m}(x)={\rm exp}(2\xi x)\,{}_{s}U_{\ell m}(x).
\label{eq:sphUtoV}
\end{eqnarray}

Substituting Eqs.~(\ref{eq:SpheU}) and (\ref{eq:SpheV}) into 
Eq.~(\ref{eq:Sph_x}), ${}_{s}U_{\ell m}(x)$ and ${}_{s}V_{\ell m}(x)$, 
respectively, satisfy the differential equations as
\begin{align}
\label{eq:proto-Jacobi}
&(1-x^{2})\,{}_{s}U_{\ell m}''(x)+\left[\beta-\alpha-(2+\alpha+\beta)x\right]\,{}_{s}U_{\ell m}'(x)\cr
&\quad +\left[\,{}_{s}E_{\ell m}(\xi)-\frac{\alpha+\beta}{2}\left(\frac{\alpha+\beta}{2}+1\right)\right]\,{}_{s}U_{\ell m}(x)\cr
=&\xi\left[-2(1-x^{2})\,{}_{s}U_{\ell m}'(x)+(\alpha+\beta+2s+2)x\,{}_{s}U_{\ell m}(x)\right.\cr
&\quad\left. -(\xi+\beta-\alpha)\,{}_{s}U_{\ell m}(x)\right],
\end{align}
and 
\begin{align}
\label{eq:proto-JacobiV}
&(1-x^{2})\,{}_{s}V_{\ell m}''(x)+\left[\beta-\alpha-(2+\alpha+\beta)x\right]\,
{}_{s}V_{\ell m}'(x)\cr
&\quad +\left[\,_{s}E_{\ell m}(\xi)-\frac{\alpha+\beta}{2}\left(\frac{\alpha+\beta}{2}+1\right)\right]\,{}_{s}V_{\ell m}(x)\cr
=&\xi\left[2(1-x^{2})\,{}_{s}V_{\ell m}'(x)-(\alpha+\beta-2s+2)x\,{}_{s}V_{\ell m}(x)\right.\cr
&\quad \left. -(\xi-\beta+\alpha)\,{}_{s}V_{\ell m}(x)\right].
\end{align}

When $\xi=0$, Eqs.~(\ref{eq:proto-Jacobi}) and (\ref{eq:proto-JacobiV}) 
reduce to the differential equation for Jacobi polynomials 
$P_{n}^{(\alpha,\beta)}(x)$ 
\begin{align}
& (1-x^{2})\,P_{n}^{(\alpha,\beta)}{}^{''}(x)
+\left[\beta-\alpha-(\alpha+\beta+2)x\right]\,P_{n}^{(\alpha,\beta)}{}^{'}(x)\cr
&\quad +n(n+\alpha+\beta+1)\,P_{n}^{(\alpha,\beta)}(x)=0,
\label{eq:Jacobi}
\end{align}
provided the eigenvalue $_{s}E_{\ell m}(\xi)$ in Eqs.~(\ref{eq:proto-Jacobi}) 
and (\ref{eq:proto-JacobiV}) becomes $\ell(\ell+1)$, 
where $\ell=n+(\alpha+\beta)/2=n+{\rm max}(\mid m\mid ,\mid s\mid )$. 
Here the Jacobi polynomials are defined by the Rodrigue's formula 
\begin{equation}
P_{n}^{(\alpha,\beta)}(x)=\frac{(-1)^{n}}{2^{n}\,n!}(1-x)^{-\alpha}(1+x)^{-\beta}\left(\frac{d}{dx}\right)^{n}\left[(1-x)^{\alpha+n}(1+x)^{\beta+n}\right].
\end{equation}

If we expand ${}_{s}U_{\ell m}(x)$ and ${}_{s}V_{\ell m}(x)$ as 
infinite series of Jacobi polynomials, 
\begin{equation}
{}_{s}U_{\ell m}(x)=\sum_{n=0}^{\infty}\,{}_{s}A_{\ell m}^{(n)}(\xi)\,P_{n}^{(\alpha,\beta)}(x),\label{eq:Jacobi-series}
\end{equation}
and 
\begin{equation}
{}_{s}V_{\ell m}(x)=\sum_{n=0}^{\infty}\,{}_{s}B_{\ell m}^{(n)}(\xi)\,P_{n}^{(\alpha,\beta)}(x),\label{eq:Jacobi-series2}
\end{equation}
we obtain three-term recurrence relations for the expansion coefficients 
${}_{s}A_{\ell m}^{(n)}(\xi)$ and ${}_{s}B_{\ell m}^{(n)}(\xi)$, respectively, as 
\begin{align}
\a^{(0)}\,{}_{s}A_{\ell m}^{(1)}(\xi)+\b^{(0)}\,{}_{s}A_{\ell m}^{(0)}(\xi)&=0, \cr
\a^{(n)}\,{}_{s}A_{\ell m}^{(n+1)}(\xi)+\b^{(n)}\,{}_{s}A_{\ell m}^{(n)}(\xi)+\c^{(n)}\,{}_{s}A_{\ell m}^{(n-1)}(\xi)&=0, \, (n\ge 1),
\label{eq:3termElm}
\end{align}
with
\begin{align}
\a^{(n)}=&\frac{4\xi(n+\alpha+1)(n+\beta+1)(n+(\alpha+\beta)/2+1-s)}{(2n+\alpha+\beta+2)(2n+\alpha+\beta+3)},\cr
\b^{(n)}=&\,_{s}E_{\ell m}(\xi)+\xi^2-\left(n+\frac{\alpha+\beta}{2}\right)\left(n+\frac{\alpha+\beta}{2}+1\right)
+\frac{2\xi s(\alpha-\beta)(\alpha+\beta)}{(2n+\alpha+\beta)(2n+\alpha+\beta+2)},\cr
\c^{(n)}=&-\frac{4\xi n(n+\alpha+\beta)(n+(\alpha+\beta)/2+s)}{(2n+\alpha+\beta-1)(2n+\alpha+\beta)},
\end{align}
and 
\begin{align}
\tilde{\a}^{(0)}\,{}_{s}B_{\ell m}^{(1)}(\xi)+\tilde{\b}^{(0)}\,{}_{s}B_{\ell m}^{(0)}(\xi)&=0, \cr
\tilde{\a}^{(n)}\,{}_{s}B_{\ell m}^{(n+1)}(\xi)+\tilde{\b}^{(n)}\,{}_{s}B_{\ell m}^{(n)}
(\xi)+\tilde{\c}^{(n)}\,{}_{s}B_{\ell m}^{(n-1)}(\xi)&=0, \quad (n\ge 1)
\label{eq:3termElm2}
\end{align}
with
\begin{align}
\tilde{\a}^{(n)}=&-\frac{4\xi(n+\alpha+1)(n+\beta+1)(n+(\alpha+\beta)/2+1+s)}{(2n+\alpha+\beta+2)
(2n+\alpha+\beta+3)},\cr
\tilde{\b}^{(n)}=&\,_{s}E_{\ell m}(\xi)+\xi^2-\left(n+\frac{\alpha+\beta}{2}\right)\left(n+\frac{\alpha+\beta}{2}+1\right)
+\frac{2\xi s(\alpha-\beta)(\alpha+\beta)}{(2n+\alpha+\beta)(2n+\alpha+\beta+2)},\cr
\tilde{\c}^{(n)}=&\frac{4\xi n(n+\alpha+\beta)(n+(\alpha+\beta)/2-s)}{(2n+\alpha+\beta-1)(2n+\alpha+\beta)}.
\end{align}
Note that, for deriving Eq.~(\ref{eq:3termElm}) and Eq.~(\ref{eq:3termElm2}), 
we use recurrence relations for Jacobi polynomials~\cite{Fackerell}. 

From the behavior of the three-term recurrence relation 
Eq.~(\ref{eq:3termElm}) for sufficiently large $n$, 
there may be two independent solutions in Eq.~(\ref{eq:3termElm}) as 
\begin{align}
&A_{(1)}^{(n)}\sim \frac{{\rm const.}(-\xi)^n}
{\Gamma(n+(\alpha+\beta+3)/2-s)}, \label{eq:AlmMin}\\
&A_{(2)}^{(n)}\sim {\rm const.}\xi^n \Gamma(n+(\alpha+\beta+1)/2+s).
\label{eq:AlmDom}
\end{align}
According to the theory of three-term recurrence relations~\cite{Gautschi}, 
$A_{(1)}^{(n)}$ is a minimal solution and $A_{(2)}^{(n)}$ is 
a dominant solution since $\lim_{n\rightarrow \infty}A_{(1)}^{(n)}/A_{(2)}^{(n)}=0$. 
The series Eq.~(\ref{eq:Jacobi-series}) computed from the dominant solution 
$A_{(2)}^{(n)}$ diverges for all values of $x$ since $A_{(2)}^{(n)}$ increases 
with $n$, while the series Eq.~(\ref{eq:Jacobi-series}) computed from 
the minimal solution $A_{(1)}^{(n)}$ converges uniformly. 
Thus, we have to choose $A_{(1)}^{(n)}$ for the series expansion
Eq.~(\ref{eq:Jacobi-series}) to obtain a solution that converges uniformly. 
This choice of $A_{(1)}^{(n)}$ requires that the eigenvalue ${}_{s}E_{\ell m}(\xi)$
satisfies a certain transcendental equation, which is expressed
in terms of continued fractions. 

In order to obtain the equation that determines the eigenvalue 
${}_{s}E_{\ell m}(\xi)$, it is convenient to introduce the following quantities: 
\begin{equation}
R_n\equiv {A_{(1)}^{(n)}\over A_{(1)}^{n-1}},\quad
L_n\equiv {A_{(1)}^{(n)}\over A_{(1)}^{n+1}}.
\end{equation}
Using the three-term recurrence relation Eq.~(\ref{eq:3termElm}) 
we can express $R_n$ as an infinite continued fraction, 
\begin{equation}
R_n=-{\gamma^{(n)}\over {\beta^{(n)}+\alpha^{(n)} R_{n+1}}}=
-{\gamma^{(n)}\over \beta^{(n)}-}
{\alpha^{(n)}\gamma^{(n+1)}\over \beta^{(n+1)}-}
{\alpha^{(n+1)}\gamma^{(n+2)}\over \beta^{(n+2)}-}\cdots, 
\label{eq:RncontElm}
\end{equation}
and $L_n$ as a finite continued fraction, 
\begin{eqnarray}
L_n=-{\alpha^{(n)}\over {\beta^{(n)}+\gamma^{(n)} L_{n-1}}}=
-{\alpha^{(n)}\over \beta^{(n)}-}\,
{\alpha^{(n-1)}\gamma^{(n)}\over \beta^{(n-1)}-}\,
{\alpha^{(n-2)}\gamma^{(n-1)}\over \beta^{(n-2)}-}\cdots
{\alpha^{(1)}\gamma^{(2)}\over \beta^{(1)}-}\,
{\alpha^{(0)}\gamma^{(1)}\over \beta^{(0)}}.
\label{eq:LncontElm}
\end{eqnarray}
The expression for $R_n$ is valid if this infinite continued fraction 
converges. Noting the properties of the three-term recurrence relations 
(see p.~35 in Ref.~\cite{Gautschi}), it can be proved that 
the continued fraction Eq.~(\ref{eq:RncontElm}) converges 
if the eigenvalue $\,_{s}E_{\ell m}(\xi)$ is finite. 

We obtain the equation to determine the eigenvalue ${}_{s}E_{\ell m}(\xi)$ 
dividing Eq.~(\ref{eq:3termElm}) by the expansion coefficients 
${}_{s}A_{\ell m}^{(n)}(\xi)$
\begin{eqnarray}
\b^{(n)}+\a^{(n)}R_{n+1}+\c^{(n)}L_{n-1}=0, 
\label{eq:determine_elm}
\end{eqnarray}
where $R_{n+1}$ and $L_{n-1}$ are defined by the continued fractions 
Eqs.~(\ref{eq:RncontElm}) and (\ref{eq:LncontElm}), which are convergent
for finite values of $\,_{s}E_{\ell m}(\xi)$. 
There are many roots in Eq.~(\ref{eq:determine_elm}) 
for given $n,m,s$ and $\xi$. These roots are associated 
with the same $m,s$, and $\xi$ but with different $\ell$. 
In order to find the root for a given $\ell$, it is useful to choose 
$n=n_{\ell}\equiv \ell-(\a+\b)/2$ in Eq.~(\ref{eq:determine_elm}) 
since in the limit $\xi\rightarrow 0$ 
all the terms in Eq.~(\ref{eq:determine_elm}) become $O(\xi^2)$. 
This means that the choice naturally gives the leading term of 
a series expansion of the eigenvalue ${}_{s}E_{\ell m}(\xi)$ 
in terms of $\xi$ as $\ell(\ell+1)$. 

When $\mid \xi\mid$ is not large, we can obtain the analytic expression of 
$_{s}E_{\ell m}(\xi)$ in a series of $\xi$ as 
\begin{eqnarray}
_{s}E_{\ell m}(\xi) = \ell (\ell +1) -\frac{2 s^2 m}{\ell (\ell +1)} \xi 
	+ \left[H(\ell+1)-H(\ell)-1\right]\xi^2 +O(\xi^3),
\end{eqnarray}
where 
\begin{eqnarray}
H(\ell)=\frac{2(\ell^2-m^2)(\ell^2-s^2)^2}{(2\ell-1)\ell^3(2\ell+1)}.
\end{eqnarray}
For the numerical calculation to determine $_{s}E_{\ell m}(\xi)$, 
one can use the analytic expression of $_{s}E_{\ell m}(\xi)$ above 
as an initial value to find the root in Eq.~(\ref{eq:determine_elm}). 

Once we obtain the eigenvalue, 
using Eqs.~(\ref{eq:RncontElm}) and (\ref{eq:LncontElm}) 
we can compute all the coefficients ${}_{s}A_{\ell m}^{(n)}(\xi)$ from 
${}_{s}A_{\ell m}^{(\tilde{n})}(\xi)$ for a given $\tilde{n}$. 
The coefficient for $n=n_\ell=l-(\alpha+\beta)/2$ is usually 
the largest term. The ratio of other terms to the dominant term, i.e. 
${}_{s}A_{\ell m}^{(n)}(\xi)/{}_{s}A_{\ell m}^{(n_\ell)}(\xi)$, 
can be determined using Eqs.~(\ref{eq:RncontElm}) and (\ref{eq:LncontElm}) 
for $0<n<n_\ell$ and $n>n_\ell$, respectively. 

We can also deal with the coefficients ${}_{s}B_{\ell m}^{(n)}(\xi)$ 
in a similar way. Noting $\tilde{\beta}^{(n)}=\beta^{(n)}$ and
$\tilde{\alpha}^{(n)}\tilde{\gamma}^{(n+1)}=\alpha^{(n)}\gamma^{(n+1)}$, 
we obtain the same equation (\ref{eq:determine_elm})
for ${}_{s}A_{\ell m}^{(n)}(\xi)$ to determine the eigenvalue $_{s}E_{\ell m}(\xi)$. 
Thus, the minimal solution of the three-term recurrence relation
for ${}_{s}B_{\ell m}^{(n)}(\xi)$, Eq.~(\ref{eq:3termElm2}), should give 
the same eigenvalue $_{s}E_{\ell m}(\xi)$ for ${}_{s}A_{\ell m}^{(n)}(\xi)$. 
Supposing that $\{B_{(1)}^{(n)}\}$ is the minimal solution, we have
\begin{align}
\frac{B_{(1)}^{(n)}}{B_{(1)}^{(n-1)}}&=
-{\tilde{\gamma}^{(n)}\over \tilde{\beta}^{(n)}-}\,
{\tilde{\alpha}^{(n)}\tilde{\gamma}^{(n+1)}\over \tilde{\beta}^{(n+1)}-}\,
{\tilde{\alpha}^{(n+1)}\tilde{\gamma}^{(n+2)}\over \tilde{\beta}^{(n+2)}-}\cdots , 
\label{eq:RncontElm2}
\\
\frac{B_{(1)}^{(n)}}{B_{(1)}^{(n+1)}}&=
-{\tilde{\alpha}^{(n)}\over \tilde{\beta}^{(n)}-}\,
{\tilde{\alpha}^{(n-1)}\tilde{\gamma}^{(n)}\over \tilde{\beta}^{(n-1)}-}\,
{\tilde{\alpha}^{(n-2)}\tilde{\gamma}^{(n-1)}\over \tilde{\beta}^{(n-2)}-}\cdots
{\tilde{\alpha}^{(1)}\tilde{\gamma}^{(2)}\over \tilde{\beta}^{(1)}-}\,
{\tilde{\alpha}^{(0)}\tilde{\gamma}^{(1)}\over \tilde{\beta}^{(0)}}.
\label{eq:LncontElm2}
\end{align}
From these equations, we can determine the ratios of 
all the coefficients, 
${}_{s}B_{\ell m}^{(n)}(\xi)/{}_{s}B_{\ell m}^{(n_\ell)}(\xi)$. 

Now we come to the problem of the normalization of the two unknown 
coefficients $A_{(1)}^{(n_{\ell})}$ and $B_{(1)}^{(n_{\ell})}$.
Since Eq.~(\ref{eq:sphUtoV}) must be satisfied for any value of $x$, 
we may substitute $x=1$ in Eq.~(\ref{eq:sphUtoV}) to obtain
\begin{eqnarray}
\,{}_{s}B_{\ell m}^{(n_{\ell})}(\xi)\sum_{n=0}^{\infty}\frac{\,{}_{s}B_{\ell m}^{(n)}(\xi)}
{\,{}_{s}B_{\ell m}^{(n_{\ell})}(\xi)}\binom{n+\a}{n}=
{\rm exp}(2\xi)\,{}_{s}A_{\ell m}^{(n_{\ell})}(\xi)\sum_{n=0}^{\infty}
\frac{\,{}_{s}A_{\ell m}^{(n)}(\xi)}{\,{}_{s}A_{\ell m}^{(n_{\ell})}(\xi)}\binom{n+\a}{n}.
\label{eq:normalization1}
\end{eqnarray}
From Eq.~(\ref{eq:normalization1}), we obtain an equation for 
$B_{(1)}^{(n_{\ell})}/A_{(1)}^{(n_{\ell})}$, i.e. 
${}_{s}B_{\ell m}^{(n_{\ell})}(\xi)/{}_{s}A_{\ell m}^{(n_{\ell})}(\xi)$,
if neither of the infinite series vanishes. 
An equation for ${}_{s}A_{\ell m}^{(n_{\ell})}(\xi)\,{}_{s}B_{\ell m}^{(n_{\ell})}(\xi)$ 
is derived from the orthogonality condition of the spin-weighted 
spheroidal harmonics Eq.~(\ref{eq:normal_S}): 
\begin{eqnarray}
\int_{-1}^{1} d x\left(\frac{1-x}{2}\right)^{\a}\left(\frac{1+x}{2}\right)^{\b}\sum_{n_{1}=0}^{\infty}\,{}_{s}A_{\ell m}^{(n_{1})}(\xi)P_{n_{1}}^{(\a,\b)}(x)\sum_{n_{2}=0}^{\infty}\,{}_{s}B_{\ell m}^{(n_{2})}(\xi)P_{n_{2}}^{(\a,\b)}(x)=1.
\label{eq:AlmBlm}
\end{eqnarray}
Using the orthogonality condition of the Jacobi polynomials, we have 
\begin{align}
&\int_{-1}^{1} d x\left(\frac{1-x}{2}\right)^{\a}\left(\frac{1+x}{2}\right)^{\b}P_{n_{1}}^{(\a,\b)}(x)P_{n_{2}}^{(\a,\b)}(x)\cr
&\quad=\frac{2\, \Gamma(n+\a+1)\Gamma(n+\b+1)\delta_{n_{1},n_{2}}}{(2n+\a+\b+1)\Gamma(n+1)\Gamma(n+\a+\b+1)}.
\end{align}
Then, Eq. (\ref{eq:AlmBlm}) reduces to 
\begin{align}
&\sum_{n=0}^{\infty}\left[\frac{\,{}_{s}A_{\ell m}^{(n)}(\xi)}{\,{}_{s}A_{\ell m}^{(n_{\ell})}(\xi)}\right]
\left[\frac{\,{}_{s}B_{\ell m}^{(n)}(\xi)}{\,{}_{s}B_{\ell m}^{(n_{\ell})}(\xi)}\right]
\frac{2\, \Gamma(n+\a+1)\Gamma(n+\b+1)}{(2n+\a+\b+1)\Gamma(n+1)
\Gamma(n+\a+\b+1)}
\cr
&\quad=\frac{1}{\,{}_{s}A_{\ell m}^{(n_{\ell})}(\xi)\,{}_{s}B_{\ell m}^{(n_{\ell})}(\xi)}.
\label{eq:normalization2}
\end{align}
From Eqs.~(\ref{eq:normalization1}) and (\ref{eq:normalization2}). 
we can obtain the squares of ${}_{s}A_{\ell m}^{(n_{\ell})}(\xi)$ and 
${}_{s}B_{\ell m}^{(n_{\ell})}(\xi)$. The final determination of 
the signs of ${}_{s}A_{\ell m}^{(n_{\ell})}(\xi)$ and ${}_{s}B_{\ell m}^{(n_{\ell})}(\xi)$ 
is made by the requirement that $_{s}S_{\ell m}^{a\omega}(x)$ reduces to 
the spin-weighted spherical harmonics in the limit $\xi\rightarrow 0$. 
\section{Homogeneous solutions of the radial Teukolsky equation}
\label{sec:PN_MST}
In this paper, we use a formalism developed by Mano, Suzuki, and Takasugi 
(MST) to compute the homogeneous solutions of 
the radial Teukolsky equation~\cite{MST,MSTR}. 
In the formalism, analytic expressions of homogeneous solutions are given 
using two kinds of series expansions in terms of 
hypergeometric functions and Coulomb wave functions, 
which are, respectively, convergent at the horizon and infinity. 
One can obtain analytic expressions of the asymptotic amplitudes 
of the homogeneous solutions by analytic matching of 
the two kinds of solutions in the overlapping region of convergence.
Compared to numerical integration methods to solve the Teukolsky equation, 
the formalism is quite powerful for very accurate numerical calculations of 
gravitational waves~\cite{FT1,FT2,FHT,Shah_20PN}. 
The formalism is also very powerful for the performance of 
post-Newtonian expansions of gravitational waves at higher orders 
since the series expansion of homogeneous solutions is closely related to 
the low-frequency expansion. 
Applying the formalism to the post-Newtonian approximation 
in black hole perturbation theory, 
the energy flux going down the horizon was calculated 
up to 6.5PN for a particle in a circular and equatorial orbit 
around a Kerr black hole~\cite{TMT} and the 2.5PN energy flux to infinity 
was computed for a particle in a slightly eccentric and inclined orbit 
around the Kerr black hole~\cite{STHGN,Ganz}. 
More recently, we applied the formalism to obtain 
the 5.5PN waveforms for a particle in a circular orbit 
around a Schwarzschild black hole~\cite{FI2010} and the 4PN waveforms 
for a particle in a circular and equatorial orbit 
around the Kerr black hole~\cite{PBFRT}, 
which, respectively, confirmed the 5.5PN energy flux in Ref.~\cite{TTS} and 
the 4PN energy flux in Ref.~\cite{TSTS}. 
In Refs.~\cite{14PN,22PN}, we extended the formalism to obtain very high PN 
expressions for the energy flux to infinity for the particle 
in a circular orbit around the Schwarzschild black hole. 
For more details of the formalism, 
we refer the reader to a recent review, Ref.~\cite{ST}. 

In the MST formalism, one can expand a homogeneous solution of 
the radial Teukolsky equation in a series of Coulomb wave functions as 
\begin{equation} 
R_{{\rm C}}^{\n}={\hat{z}}^{-1-s}\left(1-{\e \kappa \over{{\hat{z}}}}\right)^{-s-i(\e+\t)/2}\,\displaystyle\sum_{n=-\infty}^{\infty}(-i)^N\frac{(\n+1+s-i\e)_n}{(\n+1-s+i\e)_n} a_n^{\n} F_{n+\n}(-is-\e,\hat{z}),
\label{eq:Rc}
\end{equation}
where $\hat{z}=\z (r- r_-)$, $\t=(\e-m\,q)/\kappa$, 
$(a)_{n}=\Gamma(a+n)/\Gamma(a)$ and $F_{N}(\eta,z)$ 
is a Coulomb wave function defined by 
\begin{eqnarray}
F_{N}(\eta,\hat{z})=e^{-i\hat{z}}2^{N}\hat{z}^{N+1}\frac{\G(N+1-i\eta)}{\G(2N+2)}\Phi(N+1-i\eta,
2N+2;2i\hat{z}), 
\label{eq:coulomb}
\end{eqnarray}
where $\Phi(\a,\b;z)$ is the confluent hypergeometric function, 
regular at $z=0$ (see Sec. 13 in Ref.~\cite{handbook}). 
Note that the so-called renormalized angular momentum $\nu$ is introduced 
in the homogeneous solution in a series of Coulomb wave functions, 
Eq.~(\ref{eq:Rc}). $\nu$ is a generalization of $\ell$, which has a property 
such that $\n\rightarrow\ell$ as $\epsilon\rightarrow 0$, and is 
determined through conditions that the series of Coulomb wave functions, 
Eq.~(\ref{eq:Rc}), converges and 
actually represents a homogeneous solution of the radial Teukolsky equation. 

Substituting the homogeneous solution, Eq.~(\ref{eq:Rc}), into 
the radial Teukolsky equation (\ref{eq:Teu}) with $T_{\ell m\omega}=0$, 
one obtains the following three-term recurrence relation 
for the expansion coefficients $a_{n}^{\n}$: 
\begin{eqnarray}
\alpha_n^\nu a_{n+1}^{\n}+\beta_n^{\nu} a_{n}^{\n}+\gamma_n^\nu a_{n-1}^{\n}=0,
\label{eq:3term}
\end{eqnarray}
where
\begin{subequations}
\begin{align}
\a_n^\n&={i\e \kappa (n+\n+1+s+i\e)(n+\n+1+s-i\e)(n+\n+1+i\t)
\over{(n+\n+1)(2n+2\n+3)}},\\
\b_n^\n&=-\lambda-s(s+1)+(n+\n)(n+\n+1)+\e^2+\e(\e-mq)
+{\e (\e-mq)(s^2+\e^2) \over{(n+\n)(n+\n+1)}},\\
\c_n^\n&=-{i\e \kappa (n+\n-s+i\e)(n+\n-s-i\e)(n+\n-i\t)
\over{(n+\n)(2n+2\n-1)}}.
\end{align}
\label{eq:3term_abc}
\end{subequations}

The series of Coulomb wave functions, Eq.~(\ref{eq:Rc}), converges 
and represents a homogeneous solution 
of the radial Teukolsky equation if $\nu$ satisfies the following equation:
\begin{eqnarray}
R_{n}^\nu\,L_{n-1}^\nu=1,
\label{eq:consistency}
\end{eqnarray}
where $R_{n}^\nu$ and $L_{n}^\nu$ are defined in terms of 
infinite continued fractions, 
\begin{eqnarray}
R_n^\nu &\equiv& {a_{n}^{\n}\over a_{n-1}^{\n}} =
-{\gamma_n^\nu\over {\beta_n^\nu+\alpha_n^\nu R_{n+1}^\nu}}
=-{\gamma_{n}^\nu\over \beta_{n}^\nu-}\,
{\alpha_{n}^\n\gamma_{n+1}^\nu\over \beta_{n+1}^\nu-}\,
{\alpha_{n+1}^\n\gamma_{n+2}^\nu\over \beta_{n+2}^\nu-}\cdots,
\label{eq:Rncont}\\
L_n^\nu&\equiv& {a_{n}^{\n}\over a_{n+1}^{\n}} = 
-{\alpha_n^\nu\over {\beta_n^\nu+\gamma_n^\nu L_{n-1}^\nu}}
=-{\alpha_{n}^\nu\over \beta_{n}^\nu-}\,
{\alpha_{n-1}^\n\gamma_{n}^\nu\over \beta_{n-1}^\nu-}\,
{\alpha_{n-2}^\n\gamma_{n-1}^\nu\over \beta_{n-2}^\nu-}\cdots, 
\label{eq:Lncont}
\end{eqnarray}
which can be derived from the three-term recurrence relation, 
Eq.~(\ref{eq:3term}). 
Observe that one can obtain two kinds of expansion coefficients, $a_{n}^\nu$, 
from two kinds of the continued fractions, $R_{n}^\nu$ and $L_{n}^\nu$. 
If $\nu$ is chosen to satisfy Eq.~(\ref{eq:consistency}) for a certain $n$, 
the two kinds of the expansion coefficients coincide and 
the series of Coulomb wave functions, Eq.~(\ref{eq:Rc}), 
converges for $r>r_{+}$.

Since $\a_{-n}^{-\n-1}=\c_{n}^{\n}$ and $\b_{-n}^{-\n-1}=\b_{n}^{\n}$
in Eq.~(\ref{eq:3term_abc}), one finds that $a_{-n}^{-\n-1}$ satisfies 
the same recurrence relation Eq.~(\ref{eq:3term}) as $a_{n}^{\n}$. 
Then it can be shown that $R_{{\rm C}}^{-\n-1}$ is also a homogeneous solution 
of the Teukolsky equation, which converges for $r>r_{+}$.

Matching the solution in a series of Coulomb wave functions, 
which converges for $r>r_{+}$, with the one in a series of hypergeometric 
functions, which converges for $r<\infty$, 
one can obtain the incoming solution of the radial Teukolsky equation, 
$R_{\ell m\omega}^{{\rm in}}$, which converges in the entire region as 
\begin{eqnarray}
R_{\ell m\omega}^{{\rm in}}=K_{\n}R_{{\rm C}}^{\n}+K_{-\n-1}R_{{\rm C}}^{-\n-1}, 
\label{eq:secondRin}
\end{eqnarray}
where 
\begin{eqnarray}
K_{\n}&=&\frac{e^{i\e\kappa}(2\e\kappa)^{s-\n-N}2^{-s}i^{N}\Gamma(1-s-i\e-i\tau)\Gamma(N+2\n+2)}
{\Gamma(N+\n+1-s+i\e)\Gamma(N+\n+1+i\tau)\Gamma(N+\n+1+s+i\e)}\cr
&&\times\left(\sum_{n=N}^{\infty}(-1)^{n}\frac{\Gamma(n+N+2\n+1)}{(n-N)!}
\frac{\Gamma(n+\n+1+s+i\e)\Gamma(n+\n+1+i\tau)}
{\Gamma(n+\n+1-s-i\e)\Gamma(n+\n+1-i\tau)}a_{n}^{\n}\right)\cr
&&\times\left(\sum_{n=-\infty}^{N}\frac{(-1)^{n}}{(N-n)!(N+2\n+2)_{n}}
\frac{(\n+1+s-i\e)_{n}}{(\n+1-s+i\e)_{n}}a_{n}^{\n}\right)^{-1},
\end{eqnarray}
and $N$ is an arbitrary integer. The factor $K_\n$ is a constant to match 
the solutions in the overlap region of convergence, 
and is independent of the choice of $N$.

By comparing $R_{\ell m\omega}^{{\rm in}}$ in Eq.~(\ref{eq:bc_rin_rup}) 
to Eq.~(\ref{eq:secondRin}) in the limit $r^{*}\rightarrow \pm\infty$, 
one derives analytic expressions for the asymptotic amplitudes 
$B^{{\rm trans}}_{\ell m\omega}$, $B^{{\rm inc}}_{\ell m\omega}$, and 
$B^{{\rm ref}}_{\ell m\omega}$ 
in Eq.~(\ref{eq:bc_rin_rup}) as
\begin{subequations}
\begin{eqnarray}
B^{{\rm trans}}_{\ell m\omega}&=&\left(\frac{\e\kappa}{\omega}\right)^{2s}e^{i\kappa(\e+\t)(\frac{1}{2}+\frac{\ln\kappa}{1+\kappa})}\sum_{n=-\infty}^{\infty}a_{n}^{\n},\cr
B^{{\rm inc}}_{\ell m\omega}&=&\omega^{-1}\left[K_{\n}-ie^{-i\pi\n}\frac{\sin\pi(\n-s+i\e)}{\sin\pi(\n+s-i\e)}K_{-\n-1}\right]
A_{+}^{\n}e^{-i\e(\ln\e-\frac{1-\kappa}{2})},\cr
B^{{\rm ref}}_{\ell m\omega}&=&\omega^{-1-2s}[K_{\n}+ie^{i\pi\n}K_{-\n-1}]A_{-}^{\n}e^{i\e(\ln\e-\frac{1-\kappa}{2})},
\end{eqnarray}
\label{eq:asymp_amp}
\end{subequations}
where
\begin{subequations}
\begin{eqnarray}
A_{+}^{\n}&=&2^{-1+s-i\e}e^{-\frac{\pi\e}{2}}e^{\frac{\pi}{2}i(\n+1-s)}
\frac{\G(\n+1-s+i\e)}{\G(\n+1+s-i\e)}\sum_{n=-\infty}^{+\infty}a_{n}^{\n},
\\
A_{-}^{\n}&=&2^{-1-s+i\e}e^{-\frac{\pi\e}{2}}e^{\frac{-\pi}{2}i(\n+1+s)}
\sum_{n=-\infty}^{+\infty}(-1)^{n}\frac{(\n+1+s-i\e)_{n}}{(\n+1-s+i\e)_{n}}
a_{n}^{\n}. 
\end{eqnarray}
\end{subequations}
For obtaining Eq.~(\ref{eq:asymp_amp}), it is useful to note that 
the asymptotic form of $r^*$ in the limit $r^*\rightarrow\pm\infty$ takes 
\begin{subequations}
\begin{eqnarray}
\omega r^* &\rightarrow& \hat{z}+\epsilon\ln\hat{z}-\epsilon\ln\epsilon \quad {\rm for} \quad r\rightarrow\infty, \\ 
kr^* &\rightarrow& \frac{\epsilon+\tau}{2}\,\ln\left(\frac{r_+-r}{2M\kappa}\right) + \kappa\,\frac{\epsilon+\tau}{2}+\frac{\kappa\,(\epsilon+\tau)}{1+\kappa}\ln\kappa \quad {\rm for} \quad r\rightarrow r_+. 
\end{eqnarray}
\label{eq:r*_asym}
\end{subequations}

As for the other homogeneous solution $R_{\ell m\omega}^{{\rm up}}$ 
in Eq.~(\ref{eq:bc_rin_rup}), we decompose the homogeneous solution 
in a series of Coulomb wave functions $R_{C}^{\n}$ as 
\begin{eqnarray}
R_{C}^{\n}=R_{+}^{\n}+R_{-}^{\n},
\end{eqnarray}
where
\begin{eqnarray}
R_{+}^{\n}&=& 2^{\n}e^{-\pi \e}e^{i\pi(\n+1-s)}
\frac{\G(\n+1-s+i\e)}{\G(\n+1+s-i\e)}
e^{-iz}z^{\n+i(\e+\t)/2}(z-\e\kappa)^{-s-i(\e+\t)/2}\cr
&&\times  \sum_{n=-\infty}^{\infty}\,i^n\,
a_n^{\n}(2z)^n\Psi(n+\n+1-s+i\e,2n+2\n+2;2iz),\label{eq:R-}\\
R_{-}^{\n}&=& 2^{\n}e^{-\pi \e}e^{-i\pi(\n+1+s)}
e^{iz}z^{\n+i(\e+\t)/2}(z-\e\kappa)^{-s-i(\e+\t)/2}\sum_{n=-\infty}^{\infty}i^n\cr
&&\times  \frac{(\n+1+s-i\e)_n}{(\n+1-s+i\e)_n}
a_n^{\n}(2z)^n\Psi(n+\n+1+s-i\e,2n+2\n+2;-2iz)\;.
\label{eq:R+}
\end{eqnarray}
For the decomposition, we used an analytic property of the confluent 
hypergeometric function (see p.~259 in Ref.~\cite{HTF}): 
\begin{eqnarray}
\Phi(\a,\c;x)=\frac{\G(\c)}{\G(\c-\a)}e^{i\a\pi\sigma}\,\Psi(\a,\c;x)+
\frac{\G(\c)}{\G(\a)}e^{i\pi(\a-\c)\sigma}\,e^x\,\Psi(\c-\a,\c;-x),
\end{eqnarray}
where $\Psi$ is the irregular confluent hypergeometric function and 
$\sigma={\rm sgn}[{\rm Im}(x)]$ is assumed. 

Since $\Psi(\a,\b,x)\rightarrow x^{-\a}$ 
in the limit $\mid x\mid\rightarrow \infty$ 
(see Sec. 13 in Ref.~\cite{handbook}), one finds
\begin{eqnarray}
R_{+}^{\n}=A_{+}^{\n}z^{-1}e^{-i(z+\e\ln z)},\,\,\,
R_{-}^{\n}=A_{-}^{\n}z^{-1-2s}e^{i(z+\e\ln z)}
\,\,\,{\rm for}\,\,\,r\rightarrow\infty.
\end{eqnarray}
Noting the functions $R_{+}^{\n}$ and $R_{-}^{\n}$ 
have factors $e^{-iz}$ and $e^{iz}$, respectively, 
one finds that $R_{+}^{\n}$ ($R_{-}^{\n}$) is an incoming (outgoing) 
wave solution at infinity. 
Then the upgoing solution $R_{\ell m\omega}^{{\rm up}}$ is given by 
\begin{eqnarray}
R_{\ell m\omega}^{{\rm up}}=R_{-}^{\n}. 
\label{eq:RuptoR-}
\end{eqnarray}

Again noting the asymptotic form of $r^*$ in the limit 
$r^*\rightarrow\pm\infty$ in Eq.~(\ref{eq:r*_asym}) and 
comparing $R_{\ell m\omega}^{{\rm up}}$ in Eq.~(\ref{eq:bc_rin_rup})  
to Eq.~(\ref{eq:RuptoR-}) in the limit $r^{*}\rightarrow +\infty$, 
one finds the asymptotic amplitude $C^{{\rm trans}}_{\ell m\omega}$ as 
\begin{eqnarray}
C^{{\rm trans}}_{\ell m\omega}=\omega^{-1-2s}\,A_{-}^\nu\, e^{i(\epsilon\ln\epsilon-\frac{1-\kappa}{2}\epsilon)}.
\label{eq:Ctrans}
\end{eqnarray}

\end{document}